\documentclass[journel]{IEEEtran}
\pdfoutput=1
\usepackage{caption}
\usepackage{slashed}
\usepackage{amsmath}
\usepackage{mathrsfs}
\usepackage{times}
\usepackage{graphicx,enumerate}
\usepackage{subfig}
\usepackage{ifthen}
\usepackage{float}
\usepackage{hyperref}
\usepackage{multicol}
\usepackage{multirow}
\usepackage{nameref}
\usepackage{stfloats}
\usepackage{amsfonts}
\usepackage{amsbsy}
\usepackage{authblk}
\usepackage[compress]{cite}
\usepackage{extarrows}
\usepackage{epsfig}
\usepackage{pdfpages}
\usepackage{amssymb}
\usepackage{framed}
\usepackage{algorithm}
\usepackage{algorithmic}
\usepackage{mathtools}
\usepackage{authblk}
\usepackage{bigints}
\usepackage{color, soul, framed}
\usepackage{xcolor}
\newtheorem{prop}{Proposition}
\newtheorem{lemma}{Lemma}
\newtheorem{corollary}{Corollary}

\newtheorem{remark}{Remark}

\hypersetup{hidelinks}
\usepackage[T1]{fontenc}
\usepackage{lipsum} 
\allowdisplaybreaks 
\IEEEoverridecommandlockouts
\begin{document}
\title{(Non)-Coherent MU-MIMO Block Fading Channels with Finite Blocklength and Linear Processing}
\author{Junjuan~Feng,~\IEEEmembership{Member,~IEEE,} Hien Quoc Ngo,~\IEEEmembership{Senior Member,~IEEE,}
and Michail Matthaiou,~\IEEEmembership{Senior Member,~IEEE}
\thanks{Manuscript received 12 January 2022; revised 14 June 2022 and 09 August 2022; accepted 25 September 2022. The work of J. Feng and M. Matthaiou was supported by a research grant from the Department for the Economy Northern Ireland under the US-Ireland R\&D Partnership Programme.
The work of H. Q. Ngo was supported by the U.K. Research and Innovation Future Leaders Fellowships under Grant MR/S017666/1.
The associate editor coordinating the review of this article
and approving it for publication was M. Cenk Gursoy.  (\textit{Corresponding author: Junjuan Feng})}
\thanks{The authors are with the Centre for Wireless Innovation (CWI),
Queen's University Belfast, Belfast BT3 9DT, U.K. (e-mail: junjuan.feng@qub.ac.uk; hien.ngo@qub.ac.uk; m.matthaiou@qub.ac.uk).}
}
\markboth{IEEE TRANSACTIONS ON WIRELESS COMMUNICATIONS,~Vol.~xx, No.~x, xx~xx}%
{Shell \MakeLowercase{\textit{et al.}}: Bare Demo of IEEEtran.cls for IEEE Journals}
\maketitle
\IEEEpeerreviewmaketitle
%
\begin{abstract}
Driven by the stringent demands of future ultra reliable and low latency communication (URLLC), we provide a comprehensive study for a coherent and non-coherent  multiuser multiple-input multiple-output (MU-MIMO) uplink system in the finite blocklength regime. The independent and identically distributed (i.i.d.) Gaussian codebook is assumed for each user. To be more specific, the base station (BS) first uses two popular linear processing schemes to combine the signals transmitted from all users, namely maximum-ratio combining (MRC) and zero-forcing (ZF). Following it, the matched maximum-likelihood (ML) and mismatched nearest-neighbour (NN) decoding metric for the coherent and non-coherent cases are respectively employed at the BS. Under these conditions, the refined \textit{third-order achievable coding rate}, expressed as a function of the blocklength, average error probability, and the  third-order term  of the information density (called as the channel perturbation),
is derived. With this result in hand, a detailed performance analysis is then pursued, through which, we derive the asymptotic results of  the channel perturbation, achievable coding rate, channel capacity, and the channel dispersion. These theoretical results enable us to obtain
a number of interesting insights related to the impact of the finite blocklength:
i) in our system setting, massive MIMO helps to reduce the channel perturbation of the achievable coding rate, which can even be discarded without affecting the performance with just a small-to-moderate number of BS antennas and number of blocks;
ii) under the non-coherent case, even with massive MIMO, the channel estimation errors cannot be eliminated
unless the transmit powers in both the channel estimation and data transmission phases for each user are made inversely proportional to the square root of the number of BS antennas;
iii) in the non-coherent case and for fixed total blocklength, the scenarios with longer coherence  intervals and smaller number of blocks offer higher achievable coding rate.
\end{abstract}
%
%
\section{Introduction}
\label{introduction}
\IEEEPARstart{U}{ltra} reliable and low latency communication (URLLC)
 has been regarded as one of the key technological pillars for future wireless networks, since many applications, such as industrial automation, tactile internet, smart cities and remote surgery, etc. will require the reliable transmission of short packets \cite{zhang2020prospective}, \cite{hu2018swipt}.
 Thus, providing URLLC is a very timely exercise and has attracted increasing research attention over the past decade.  For instance, \cite{makki2016wireless} analyzed the effect of  wireless energy transmission and the information signals' length on the outage probability,  while \cite{lopez2020finite}  provided accurate analytical approximations for the error probability distribution.
Now, recall that the most prevalent used metric in physical layer wireless communication analysis is the Shannon capacity, which is the largest coding rate at which the error probability can be made arbitrarily small by choosing a sufficiently large blocklength.
Thus, it is accurate enough for the performance evaluation
of wireless systems, whenever it is not necessary to consider the latency constraints. However, with stringent latency requirements, as those in URLLC, the long codewords assumption becomes irrelevant and we should instead resort to short-codeword transmissions, which results in a non-zero error probability and significant rate loss. In this context, using Shannon capacity will not only overestimate the system performance but also offer confusing guidelines for system design in the finite blocklength regime. Taking this into account,
one of the most important metrics in URLLC, i.e., \textit{the refined maximum achievable channel coding rate} is introduced,  which incorporates the channel dispersion resulting from the finite blocklength and  is expressed as a function of the blocklength and error probability \cite{polyanskiy2010channel}.   This metric has been proved to predict the performance more accurately in the finite blocklength regime.

The work of \cite{polyanskiy2010channel} triggered considerable research interest and a plethora of research groups have extended its results to more general and complicated fading channels.
Specifically,
\cite{scarlett2017dispersion} has taken a step further to present a channel dispersion analysis under additive non-Gaussian noise and random Gaussian codebooks along with the nearest-neighbor (NN) decoding.
For  fading channels,
\cite{collins2018coherent} studied the dispersion over coherent multiple-input multiple-output (MIMO) block-fading channels, and showed that the channel dispersion approaches to a constant with an increasing number of receive antennas. Regarding the non-coherent scenario,
\cite{ostman2019short,lancho2019single,
lancho2020saddlepoint}
analyzed the maximum coding rate under  block-fading channels.
More specifically, \cite{ostman2019short} presented non-asymptotic bounds on the maximum coding rate under  non-coherent  Rician block-fading channels and quantified the optimal trade-off between the rate gain obtained from the channel diversity and rate loss resulting from fast channel dynamics and pilot overhead. Following it, \cite{lancho2019single} provided a high signal-to-noise ratio (SNR) normal approximation of the maximum coding rate, which complements the non-asymptotic bounds in \cite{ostman2019short} and provided us with a tractable formula for performance analysis.
Later, \cite{lancho2020saddlepoint} generalized the high {\rm SNR} result of \cite{lancho2019single} via saddlepoint approximations and proved that these new approximations are also accurate in the low {\rm SNR}s regime.

Before presenting the motivations and contributions of our work, it is worth recalling the importance of massive multiple-input multiple-output (MIMO) in the wireless space over the past decade, since it can provide substantial spectral efficiency while guaranteeing high reliability \cite{matthaiou2021road},\cite{durisi2016short}.
Likewise, multiuser MIMO (MU-MIMO) topologies afford an additional dimension for spatially multiplexing geographically separated users \cite{Larsson2014}. In this context, we
refer to \cite{ngo2013energy} which showed that nearly optimal performance can be obtained by employing linear processing techniques, such as maximum-ratio combining (MRC) and zero-forcing (ZF) schemes,  when the number of BS antennas is large. Surprisingly, the combination of finite blocklength systems with MU-MIMO and linear signal processing techniques has been largely overlooked in the related literature due to the existence of inter-user interference and channel estimation errors, which renders the performance analysis challenging.  Thus, this paper tries to make a first step towards bridging this gap.

To date, most works in the finite blocklength space have mainly  focused on general point-to-point networks \cite{polyanskiy2010channel,
scarlett2017dispersion,
collins2018coherent,
ostman2019short,
lancho2019single,lancho2020saddlepoint}.
 The only exception is  \cite{ostman2021urllc}, which provided a detailed  analysis of the random coding union (RCU) bound of the error probability via the
 saddlepoint approximation for both the uplink and downlink of massive MIMO system  under imperfect channel state information (CSI). Recently, there has been a surge of interest in studying the  higher-order terms of the achievable coding rate in the finite blocklength  regime \cite{tan2015third}.
However, most existing works focus on the scaling order of the third-order term along with the blocklength for point-to-point networks, except \cite{lancho2019single}, in which
a loose upper bound on the third-order term was derived.
In practice, whether the high-order terms of the achievable coding rate, e.g., the channel dispersion and  third-order term, are bounded or not with massive MIMO remains questionable. This characteristic is critical when using the RCU bound and the simpler Berry-Esseen central limit theorem (BE-CLT) as a basis.

%
For the reasons mentioned above, it is indispensable to pursue a performance analysis for MU-MIMO systems in the finite blocklength space and for fading channels. Motivated by this, the main contributions of this paper are listed as follows:
\begin{itemize}
\item
We study the MU-MIMO uplink system in the finite blocklength regime, in which a BS equipped with multiple antennas simultaneously serves multiple single-antenna users. The linear combining schemes, MRC and ZF, are applied at the BS. Both coherent and non-coherent block-fading channels with independent and identically distributed (i.i.d.) Gaussian input signals are considered.
The matched maximum-likelihood (ML) and mismatched nearest-neighbour (NN) decoding metric are applied for the coherent and non-coherent case, respectively.
Two of the fundamental performance metrics,  the achievable coding rate and average decoding error probability, are investigated.
Note that the achievable coding rate has been the primary focus in the existing literature since it is adequate to characterize the behavior of the finite blocklength \cite{scarlett2020information}.
%
\item
Closed-form results of the channel dispersion and the third-order term (the channel perturbation in this paper) of the achievable coding rate are derived for both the coherent and non-coherent cases. Note that for the channel perturbation term in the non-coherent case, a tight approximation is provided in terms of the matched Gamma distribution approximation.
%
\item Based on the RCU bound of the average error probability and further the BE-CLT, the achievable coding rate is derived under the condition that the channel perturbation is less than a given threshold. Capitalizing on this result, a closed-form expression of the BE-CLT based upper bound of the average error probability is given. By proving that the channel perturbation  is quite small (smaller than 1) with a small-to-moderate number of BS antennas and blocks, a Taylor expansion is applied on the derived achievable coding rate and a new approximation of the achievable coding rate is provided by neglecting the channel perturbation term. Following it, a Taylor expansion based approximation of the average error probability is given.
\item
    New asymptotic results with and without considering the power scaling law under massive MIMO, including the achievable coding rate,   channel capacity,  channel dispersion, and  channel perturbation, are derived. From  these results and simulations, we can obtain some important and useful insights:
\begin{enumerate}
\item In MU-MIMO systems with linear combining schemes and i.i.d Gaussian input signals, massive antenna arrays at the BS help to reduce the channel perturbation of the achievable coding rate to a small constant, whilst a large enough number of blocks pushes the channel perturbation to zero. Meanwhile, the channel dispersion of the achievable coding rate also approaches to a constant with massive antenna arrays at the BS.
\item Having a large number of antennas at the BS can help to reduce the minimum blocklength required for obtaining a certain fraction of  the channel capacity and also helps to reduce the  average error probability.
    Likewise, increasing the number of BS antennas will improve the achievable coding rate even when the channel dispersion (rate loss) caused by the finite blocklength is incorporated into the expression.
\item Different from the infinite blocklength scenario, if no power scaling laws are considered, massive MIMO cannot eliminate the channel estimation errors  for the non-coherent case in the finite blocklength space. Consequently, the corresponding  effective noise vector cannot be approximated by a Gaussian distribution. However, when the power scaling law at both the channel estimation and data transmission phases are considered, all channel estimation errors can be canceled out and both the effective noise and the output signal vector tend to be Gaussian distributed with massive MIMO.
\item For the non-coherent case, by fixing the total blocklength, the scenarios with  longer coherence intervals and smaller number of blocks are more beneficial for serving the higher achievable coding rate. On the contrary, in the coherent case, the normal approximation of the achievable coding rate is equally affected by these two parameters. Nevertheless, from a channel perturbation point of view, the system with more number of blocks and a shorter coherence interval is better.
\end{enumerate}
\end{itemize}
%

{\bf Notation:} Boldface upper and lower case letters are used to denote matrices and vectors respectively. The symbols ${\bf A}_{:,k}$ and ${\bf x}_{m}$ denote the $k$-th column of the matrix ${\bf A}$ and the $m$-th element of the vector ${\bf x}$, respectively. The superscripts $(\cdot)^\dagger$, $(\cdot)^T$ and $(\cdot)^*$ are used to denote the conjugate transpose, the transpose and the conjugate of a matrix or a vector, respectively. The notation $\|{\bf x}\|_{2}=\sqrt{{\bf x}^{\dagger}{\bf x}}$ denotes the 2-norm of the complex column vector ${\bf x}$. The notation $\mathbb{E}\{\cdot\}$ denotes the expectation of random variables and the notation $\mathcal{CN}(0, \sigma^2)$  represents the circularly-symmetric complex Gaussian distribution with zero mean and variance $\sigma^2$.
The symbol $\chi^{2}(N)$ denotes the chi-square distribution with $N$ degrees of freedom. The notation ${\rm diag}\{\cdot\}$ denotes the diagonal operator.
Finally, $f(x)=\mathcal{O}(g(x))$ means that $\lim\limits_{x\rightarrow \infty}\frac{f(x)}{g(x)}=c<\infty$.
\vspace{-3mm}
\section{System Model}\label{section_2}
We consider a multiuser uplink system consisting of $K_{u}$ single-antenna users and one BS equipped with $N_{b}$ antennas in the finite blocklength regime. As previously mentioned, a block-fading channel is considered, i.e., the channel remains constant during one coherence block and varies independently to a new realization in the next block.
 We assume that there are in total $L$ blocks and each block has $T_{c}$ symbols, such that the total blocklength for users to transmit a message is $n=LT_{c}$. The pilot length allocated for the channel estimation in each block is assumed to be $\tau_{c}$. Thus, we define the notation $\tilde{T}_{c}$, which represents the number of symbols used for data transmissions in each block,  as $\tilde{T}_{c}=T_{c}$ in the coherent case and $\tilde{T}_{c}=T_{c}-\tau_{c}$ in the non-coherent case.
The detailed data transmission process is as follows.

At block $j$, for $j=1,2,\ldots,L$, for each symbol $i\in[\tilde{T}_{c}]=\{1,2,\ldots,\tilde{T}_{c}\}$, the received signal ${\bf y}_{i}(j)\in\mathcal{C}^{N_{b}\times 1}$ at the BS is
%
\begin{align}\label{eq_20190511_1a}
{\bf y}_{i}(j)=\sum\limits^{K_{u}}_{k=1}{\bf h}_{k}(j) x_{k,i}(j)+{\bf n}_{i}(j),
\end{align}
%
where ${\bf h}_{k}(j)=[h_{k,1}(j),h_{k,2}(j),
\ldots,h_{k,N_{b}}(j)]^{T}\in\mathcal{C}^{N_{b}\times 1}$ denotes the channel realization from the $k$-th user to the BS at block  $j$ with each element drawn independently from the distribution $\mathcal{CN}(0, \gamma_{k}^2)$, in which $\gamma_{k}^2$ denotes the large-scale fading coefficient from the $k$-th user to the BS. The random variable $x_{k,i}(j)$ is the $i$-th transmit symbol of user $k$ at block $j$, and ${\bf n}_{i}(j)=[n_{i,1}(j),\ldots,n_{i,N_{b}}(j)]^{T}\in\mathcal{C}^{N_{b}\times1}$ is the additive white Gaussian noise (AWGN) vector, which has i.i.d. $\mathcal{CN}(0, \sigma^2_{n})$ elements.

From \eqref{eq_20190511_1a},  the received  signal matrix at the BS   for all symbols $[\tilde{T}_{c}]$ at  block $j$ is:
%
\begin{align}\label{eq_20190511_1}
{\bf Y}(j)=\sum\limits^{K_{u}}_{k=1}{\bf h}_{k}(j)
{\bf x}^{T}_{k}(j)+{\bf N}(j),
\end{align}
%
where ${\bf Y}(j)=[{\bf y}_{1}(j),\ldots,{\bf y}_{\tilde{T}_{c}}(j)]\in\mathcal{C}^{N_{b}\times \tilde{T}_{c}}$,  ${\bf N}(j)=[{\bf n}_{1}(j),\ldots,{\bf n}_{\tilde{T}_{c}}(j)]\in \mathcal{C}^{N_{b}\times \tilde{T}_{c}}$, and
${\bf x}_{k}(j)=[x_{k,1}(j),\ldots,x_{k,\tilde{T}_{c}}(j)]^{T}\in \mathcal{C}^{\tilde{T}_{c}\times 1}$ contains all the transmitted symbols within block $j$ of user $k$. We assume that the input signal vectors from all users $\{{\bf x}_{k}(j)\}^{K_{u}}_{k=1}$ are independent and each signal vector ${\bf x}_{k}(j)$ is drawn from the distribution $\mathcal{CN}(0, P_{k}{\bf I}_{\tilde{T}_{c}})$;\footnote{Note that three popular codebooks are the i.i.d. Gaussian input signals, spherical Gaussian codebooks (also called shell codes) in \cite{scarlett2017dispersion}, and the unitary space time modulation (USTM) in \cite{lancho2019single}, wherein the shell codes are a special case of the USTM codes. The i.i.d. Gaussian input signals assumed in this paper do not represent the optimal input distribution in the finite blocklength regime. They are capacity-achieving input distributions for the matched decoding metric (considered in the coherent case), but not the optimal distribution for the channel dispersion metric, compared with other existing input distributions, such as the shell codes, which can achieve a lower channel dispersion than the i.i.d. Gaussian input distribution. Moreover, the USTM is not capacity-achieving at low to median SNRs for non-coherent block fading channels.
Here, we use an i.i.d. Gaussian input codebook as
it is a common codebook in the literature \cite{scarlett2020information} and also amenable to mathematical manipulations later on (similar to \cite{ostman2021urllc}, \cite{hoydis2012randomc}), which facilitates the system design and optimization. Moreover, we know from \cite{scarlett2017dispersion}, \cite{nasir2021cell}, that the channel dispersion of the i.i.d. Gaussian codebook is close to that of the shell codes under a large amount of inter-user interference, which is the exact case we consider in our paper.}
 thus, we have
 %
\begin{align}\label{eq_20191219_1}
\mathbb{E}\left\{\|{\bf x}_{k}(j)\|_{2}^2\right\}=\tilde{T}_{c}P_{k}, \quad k=1,\ldots,K_{u},
\end{align}
%
which can be treated as the average transmit power for each user. Here, we point out that all the channels, input signals and noise signals are respectively block-wise independent, and also independent of each other.

Following the uplink transmission process in \cite{ostman2021urllc}, for an intended user $k$ at the block $j$, the BS will process its received signals with a linear combining vector ${\bf a}_{k}(j)\in\mathcal{C}^{N_{b}\times 1}$. By multiplying the received signal matrix ${\bf Y}(j)$ in \eqref{eq_20190511_1} with ${\bf a}^{\dagger}_{k}(j)$, we obtain
%
\begin{align}\label{eq_20191015_1}
\tilde{{\bf y}}_{k}(j)&={\bf a}^{\dagger}_{k}(j){\bf h}_{k}(j){\bf x}^{T}_{k}(j)\notag\\
&\quad+\sum\limits^{K_{u}}_{\substack{m=1\\ m\neq k}}{\bf a}^{\dagger}_{k}(j){\bf h}_{m}(j){\bf x}^{T}_{m}(j)+{\bf a}^{\dagger}_{k}(j){\bf N}(j),
\end{align}
%
where $\tilde{{\bf y}}_{k}(j)\in\mathcal{C}^{1\times \tilde{T}_{c}}$ is the processed signal row vector (after linear processing) used for decoding the symbols transmitted from user $k$.

Assume that $M_{k}$ is the maximum number of messages that are allowed to be transmitted over the fading channels by the $k$-th user. Note that as we consider the instantaneous rate, $M_{k}$ is a function of the fading coefficients.
Then, we will introduce the channel coding notations such that the targeted user $k$ transmits $M_{k}$ messages under the average decoding error probability
no larger than $0<\epsilon_{k}<1$ and the finite data transmission blocklength $\tilde{n}=L\tilde{T}_{c}$. A  $(M_{k},L,\tilde{T}_{c},\epsilon_{k})$-code consists of the following two functions:
\begin{enumerate}[1)]
\item An encoding function $f_{k}$: $[M_{k}]=\{1,\ldots, M_{k}\}\rightarrow \mathcal{C}^{\tilde{T}_{c}\times L}$: this function is used to map the message $W_{k}$, which takes values uniformly on the set $[M_{k}]$, to a codeword $({\bf X}_{k})^{L}\triangleq[{\bf x}_{k}(1),\ldots,{\bf x}_{k}(L)]$ under the power constraint in \eqref{eq_20191219_1};
\item A decoding function $g_{k}$: $\mathcal{C}^{L\times \tilde{T}_{c}}\!\!\rightarrow\!\![M_{k}]$: this function can decode the message $\!W_{k}\!$ from the received signals $\!{\small(\widetilde{{\bf Y}}_{k})^{L}\triangleq[\tilde{{\bf y}}^{T}_{k}(1),\ldots,\tilde{{\bf y}}^{T}_{k}(L)]^{T}}\!\!$ under the average error probability $\!\epsilon_{k}$, which is defined as
%
\begin{align}\label{eq_20200118_1}
\epsilon_{k}\triangleq \frac{1}{M_{k}}
\sum\limits_{w_{k}\in[M_{k}]}
\mathbb{P}\{g_{k}((\widetilde{{\bf Y}}_{k})^{L})\neq w_{k}|W_{k}=w_{k}\},
\end{align}
where the notation $\mathbb{P}\{A\}$ denotes the probability that the event "A" happens.
\end{enumerate}

In the following sections, we analyze the performance of our considered system in terms of the achievable coding rate and average error probability for two common linear combining schemes: MRC and ZF. Another common linear combining scheme is the minimum mean-square error (MMSE) scheme. However, the analysis of this scheme is rather complicated. More importantly, the performance of MRC is very close to that of MMSE at low SNRs, while the performance of ZF approaches that of MMSE at medium and high SNRs \cite{ngo2015massive}. So, the analysis of MRC and ZF can reasonably cover the analysis of MMSE.
\vspace{-2mm}
\section{Coherent channels}\label{perfect_csi}
In this section, the coherent case is considered, i.e., all $LT_{c}$ symbols are used for transmitting signals from all users and there are no resources  allocated for the pilot transmission.
\vspace{-3mm}
\subsection{MRC Scheme}\label{mrc_scheme_trans}
For coherent channels, i.e., perfect CSI at the BS,
when the MRC scheme is applied at the BS, the vector ${\bf a}_{k}(j)$ at the $j$-th block in \eqref{eq_20191015_1} is
%
\begin{align}\label{eq_20191031_1}
{\bf a}_{k,\text{co}}^{\text{mrc}}(j)
=\frac{{\bf h}_{k}(j)}{\|{\bf h}_{k}(j)\|_{2}},
\end{align}
for all $k=1,\ldots,K_{u}$ and $j=1,\ldots, L$.
By substituting \eqref{eq_20191031_1} into \eqref{eq_20191015_1}, we obtain the received signals at the BS after combining as
%
\begin{align}\label{eq_20190729_3}
\tilde{{\bf y}}^{\text{mrc}}_{k,\text{co}}(j)
&=\underbrace{\|{\bf h}_{k}(j)\|_{2}{\bf x}^{T}_{k}(j)}_{\text{Desired signal}}
+\underbrace{\sum\limits^{K_{u}}_{\substack{m=1\\ m\neq k}}\frac{{\bf h}^{\dagger}_{k}(j)}{\|{\bf h}_{k}(j)\|_{2}}
{\bf h}_{m}(j) {\bf x}^{T}_{m}(j)}_{\text{Inter-user interference}}\notag\\
&\quad+\underbrace{\frac{{\bf h}^{\dagger}_{k}(j)}{\|{\bf h}_{k}(j)\|_{2}}{\bf N}(j)}_{\text{Noise}}.
\end{align}
%
We denote by
%
\begin{align}\label{eq-2021-11-27-1}
\!\!\check{{\bf n}}_{\text{co}}^{\text{mrc}}(j)
\!\triangleq\!
\sum\limits^{K_{u}}_{\substack{m=1\\ m\neq k}}
\frac{{\bf h}^{\dagger}_{k}(j)}{\|{\bf h}_{k}(j)\|_{2}}
{\bf h}_{m}(j) {\bf x}^{T}_{m}(j)
\!+\!
\frac{{\bf h}^{\dagger}_{k}(j)}{\|{\bf h}_{k}(j)\|_{2}}{\bf N}(j),
\end{align}
the effective noise including the interference plus noise. Then, \eqref{eq_20190729_3} can be rewritten as
%
\begin{align}
\label{eq_20190729_3b}
\tilde{{\bf y}}^{\text{mrc}}_{k,\text{co}}(j)
= \|{\bf h}_{k}(j)\|_{2}{\bf x}^{T}_{k}(j) + \check{{\bf n}}_{\text{co}}^{\text{mrc}}(j),
\end{align}
%
where $\tilde{{\bf y}}^{\text{mrc}}_{k,\text{co}}(j)\in\mathcal{C}^{1\times T_{c}}$.
Under the assumptions of perfect CSI at the BS and the i.i.d.  Gaussian input signals, both the interference term  and the noise term in \eqref{eq_20190729_3} are still Gaussian distributed. So, the effective noise $\check{{\bf n}}_{\text{co}}^{\text{mrc}}(j)\in\mathcal{C}^{1\times T_{c}}$ is also a Gaussian vector with the distribution $\check{{\bf n}}_{\text{co}}^{\text{mrc}}(j)\sim \mathcal{CN}\left({\bf 0}, \sigma^{2}_{\check{{\bf n}}_{\text{co}}^{\text{mrc}}}(j){\bf I}_{T_{c}}\right)$, in which $\sigma^{2}_{\check{{\bf n}}_{\text{co}}^{\text{mrc}}}(j)$ is given by
%
\begin{align}\label{eq_20200120_5a}
\sigma^{2}_{\check{{\bf n}}_{\text{co}}^{\text{mrc}}}(j)
=\sum\limits^{K_{u}}_{\substack{m=1\\ m\neq k}}P_{m}\left|\frac{{\bf h}^{\dagger}_{k}(j)}{\|{\bf h}_{k}(j)\|_{2}}{\bf h}_{m}(j)\right|^{2}+\sigma^{2}_{n}.
\end{align}
%
So, the end-to-end channel model \eqref{eq_20190729_3b} can be considered as a point-to-point SISO channel with the i.i.d. Gaussian input signal vector ${\bf x}^{T}_{k}(j)$, the coherent fading channel  $\|{\bf h}_{k}(j)\|_{2}$ and the i.i.d. Gaussian noise $\check{{\bf n}}_{\text{co}}^{\text{mrc}}(j)$. The corresponding signal-to-interference-plus-noise-ratio ${\rm SINR}^{{\text{mrc}}}_{k,\text{co}}(j)$ for the $k$-th user
at the $j$-th block is
%
\begin{align}\label{sinr_mrc}
{\rm SINR}^{{\text{mrc}}}_{k,\text{co}}(j)
=\frac{P_{k}\|{\bf h}_{k}(j)\|^{2}_{2}}
{\sum\limits^{K_{u}}_{\substack{m=1\\ m\neq k}}P_{m}
\left|\frac{{\bf h}^{\dagger}_{k}(j)}{\|{\bf h}_{k}(j)\|_{2}}{\bf h}_{m}(j)\right|^{2}+\sigma^{2}_{n}}.
\end{align}

Hence, the matched decoding metric, i.e., the ML decoder, could be applied at the BS. To be consistent with the non-coherent case later, we introduce a coefficient $s$, i.e., for arbitrary $s>0$, namely the generalized ML decoding metric, which is given by
%
\begin{align}\label{eq_p_mrc_20210307_1}
(\hat{{\bf X}}_{k})^{L}
=\underset{\left({\bf X}_{k}\right)^{L}\in\mathcal{C}^{T_{c}\times L}}
{\arg \max} \mathrm{p}^{s}\left\{\left.\left((\tilde{{\bf Y}}_{k})_{\text{co}}^{\text{mrc}}\right)^{L}\right|
\left({\bf X}_{k}\right)^{L},\left({\bf H}_{k}\right)^{L}\right\},
\end{align}
%
where
%
\begin{align}\label{eq_20200731_1}
&\quad\mathrm{p}^{s}\left\{\left.\left((\tilde{{\bf Y}}_{k})_{\text{co}}^{\text{mrc}}\right)^{L}\right|
\left({\bf X}_{k}\right)^{L},\left({\bf H}_{k}\right)^{L}\right\}\notag\\
&=\prod\limits^{L}_{j=1}\mathrm{p}^{s}
\left\{\left.\tilde{{\bf y}}^{\text{mrc}}_{k,\text{co}}(j)\right|
{\bf x}_{k}(j),\|{\bf h}_{k}(j)\|_{2}\right\}\notag\\
&=\prod\limits^{L}_{j=1}(\pi)^{-sT_{c}}
\left[\sigma^{2}_{\check{{\bf n}}_{\text{co}}^{\text{mrc}}}(j)
\right]^{-sT_{c}}\notag\\
&\quad\quad\times\exp\left[-s
\left(\frac{1}{\sigma^{2}_{\check{{\bf n}}_{\text{co}}^{\text{mrc}}}(j)}
\right)
(\check{{\bf n}}_{\text{co}}^{\text{mrc}}(j))
(\check{{\bf n}}_{\text{co}}^{\text{mrc}}(j))^{\dagger}\right],
\end{align}
%
in which {\small$({\bf X}_{k})^{L}\in\mathcal{C}^{T_{c}\times L}$} and {\small$\left((\tilde{{\bf Y}}_{k})_{\text{co}}^{\text{mrc}}\right)^{L}\in\mathcal{C}^{L\times T_{c}}$} are defined similarly as  in Section \ref{section_2}. The channel matrix {\small$\left({\bf H}_{k}\right)^{L}
=\operatorname{diag}\left\{\|{\bf h}_{k}(1)\|_{2},\ldots,\|{\bf h}_{k}(L)\|_{2}\right\}$}.
It is well known that when $s=1$, \eqref{eq_20200731_1} coincides with the capacity-achieving output distribution for the matched ML decoding metric,
i.e., the i.i.d. Gaussian inputs are capacity-achieving for the matched decoding metric in the coherent case \cite[Section 3.8.1]{scarlett2020information}.
Thus, with the decoding metric in \eqref{eq_p_mrc_20210307_1}-\eqref{eq_20200731_1} and setting $s=1$, the following proposition can be obtained.
%
\begin{prop}\label{prop_20200103_1}
Consider perfect CSI at the BS and assume that the input signals are i.i.d. Gaussian distributed. Under the MRC scheme, define a threshold term
%
\begin{align}\label{eq_20210923_2}
{q}^{\text{mrc},\text{co}}_{k,\text{thres}}
&\triangleq
Q\left(\left(\frac{\frac{1}{L}\sum\limits^{L}_{j=1}
V_{k,\text{co}}^{\text{mrc}}(j)}
{LT_{c}}\right)^{-\frac{1}{2}}\right.\notag\\
&\left.\quad\quad\times\left(\frac{\ln M_{k}}{LT_{c}}-\frac{\ln\mu}{LT_{c}}
-\frac{1}{L}
\sum\limits^{L}_{j=1}
I_{k,\text{co}}^{\text{mrc},\text{im}}(j)\right)\right),
\end{align}
where $Q(\cdot)$ is the $Q$-function.
When the channel perturbation term ${U}^{\text{mrc}}_{k,\text{co}}$, which will be defined later, satisfies the condition  ${U}^{\text{mrc}}_{k,\text{co}}\leq {q}^{\text{mrc},\text{co}}_{k,\text{thres}}$, the
relation between the average error probability ${\epsilon}_{k,\text{co}}^{\text{mrc}}$ and the achievable coding rate $R^{\text{mrc}}_{k,\text{co}}$ (in nats per channel use)\footnote{The achievable coding rate represents the achievability bound (lower bound) of the maximum coding rate, and it does not represent the actual maximum coding rate.
Note that throughout the  paper, we will use the achievable coding rate as our performance metric.}  is given as \eqref{eq_20210305_2} shown at the top of next page.
%
\begin{figure*}[!ht]
\begin{align}
\label{eq_20210305_2}
\frac{\ln M_{k}}{LT_{c}}
\geq
R^{\text{mrc}}_{k,\text{co}}
\!\triangleq\!
\frac{1}{L}
\sum\limits^{L}_{j=1}I_{k,\text{co}}^{\text{mrc}}(j)
\!+\!\frac{\ln\mu}{LT_{c}}
\!-\!\sqrt{\frac{\frac{1}{L}\sum\limits^{L}_{j=1}
{V}_{k,\text{co}}^{\text{mrc}}(j)}{LT_{c}}}
Q^{-1}\left({\epsilon}_{k,\text{co}}^{\text{mrc}}
\right)
+\sqrt{\frac{\frac{1}{L}\sum\limits^{L}_{j=1}
{V}_{k,\text{co}}^{\text{mrc}}(j)}{LT_{c}}}
\mathcal{O}\left({U}^{\text{mrc}}_{k,\text{co}}
\right).
\end{align}
\center\hrulefill
\vspace{-3mm}
\end{figure*}
In \eqref{eq_20210923_2} and \eqref{eq_20210305_2},  the parameter $\mu$ is uniformly distributed on the interval $(0,1)$. 
The term $I_{k,\text{co}}^{\text{mrc}}(j)$ represents the channel capacity at the $j$-th block, given by
%
\begin{align}\label{eq_20200103_2}
I_{k,\text{co}}^{\text{mrc}}(j)
=\ln\left(1+{\rm SINR}^{{\text{mrc}}}_{k,\text{co}}(j)\right).
\end{align}
%
Furthermore,  $V^{\text{mrc}}_{k,\text{co}}(j)$ denotes the channel dispersion, which is used to characterize the backoff of the achievable coding rate from the capacity caused by the finite blocklength,  is
%
\begin{align}\label{eq_20200103_3}
V^{\text{mrc}}_{k,\text{co}}(j)
&=2-\frac{2}{1+{\rm SINR}^{{\text{mrc}}}_{k,\text{co}}(j)}.
\end{align}
%
The term ${U}^{\text{mrc}}_{k,\text{co}}$, which is defined in \eqref{eq_per_mrc_20210317_1} of Appendix \ref{appendix prop_20200103_1} and stems from the absolute third-order moment of the information density, is called the channel perturbation and given by,
%
\begin{align}\label{eq_u_1}
{U}^{\text{mrc}}_{k,\text{co}}
&=\frac{4c_{1}}{\sqrt{2\pi}}\frac{1}{\sqrt{L}}
\frac{1}{\sqrt{T_{c}}\Gamma\left(T_{c}+1\right)}
\Gamma\left(\frac{2T_{c}+3}{2}\right)\notag\\
&\quad\times
\left(\!\frac{1}{L}\sum\limits^{L}_{j=1}{V}_{k,\text{co}}^{\text{mrc}}(j)
\!\right)^{-\frac{3}{2}}\!\!
\left\{\frac{1}{L}\sum\limits^{L}_{j=1}
\left[V^{\text{mrc}}_{k,\text{co}}(j)
\right]^{\frac{3}{2}}\right\}.
\end{align}
\end{prop}
%
\begin{IEEEproof}
See Appendix \ref{appendix prop_20200103_1}.
\end{IEEEproof}
\begin{remark}\label{remark_mrc_per}
It should be noted that both the channel dispersion in \eqref{eq_20200103_3} and the channel perturbation in \eqref{eq_u_1} of the achievable coding rate are obtained by leveraging the capacity-achieving input distribution ($s=1$).
From Appendix \ref{appendix prop_20200103_1}, the result in Proposition \ref{prop_20200103_1} is obtained
based on the RCU bound of the average error probability in \cite[Theorem 1]{martinez2011saddlepoint} and the Berry-Esseen central limit theorem (BE-CLT) \cite{maya2016closed}.
From \eqref{eq_20200731_4} of
 Appendix \ref{appendix prop_20200103_1}, we have that when ${U}^{\text{mrc}}_{k,\text{co}}
>{q}^{\text{mrc},\text{co}}_{k,\text{thres}}$,
the largest allowable (upper bound) average error probability ${\epsilon}_{k,\text{co}}^{\text{mrc}}$  is 1. However, when ${U}^{\text{mrc}}_{k,\text{co}}\leq {q}^{\text{mrc},\text{co}}_{k,\text{thres}}$, the average error probability $\epsilon^{\text{mrc}}_{k,\text{co}}$ after using the BE-CLT is upper bounded by
%
\begin{align}\label{eq_20210917_3}
\epsilon^{\text{mrc}}_{k,\text{co}}
\leq 1-{q}^{\text{mrc},\text{co}}_{k,\text{thres}}
+{U}^{\text{mrc}}_{k,\text{co}}
\triangleq \epsilon^{\text{mrc-beclt}}_{k,\text{co}}.
\end{align}
%

In general, the desirable average error
probability $\epsilon^{\text{mrc}}_{k,\text{co}}$ should be at least smaller than $0.5$, whilst in URLLC, the average error probability
$\epsilon^{\text{mrc}}_{k,\text{co}}$ should be less than $10^{-5}$ for one transmission package \cite{he2021beamforming}. In this context, the scenario ${U}^{\text{mrc}}_{k,\text{co}}
>{q}^{\text{mrc},\text{co}}_{k,\text{thres}}$ is not desirable. Fortunately, from \eqref{eq_u_1}, after fixing the number of BS antennas,
the channel  perturbation  ${U}^{\text{mrc}}_{k,\text{co}}$ decreases with respect to $L$ according to $\mathcal{O}\left(\frac{1}{\sqrt{L}}\right)$, whilst the function ${q}^{\text{mrc},\text{co}}_{k,\text{thres}}$ in \eqref{eq_20210923_2} is non-decreasing with respect to $L$. Thus, the condition  ${U}^{\text{mrc}}_{k,\text{co}}\leq {q}^{\text{mrc},\text{im}}_{k,\text{thres}}\leq 1$ can be satisfied by increasing the number of blocks $L$. With this, by substituting \eqref{eq_20210923_2} into \eqref{eq_20210917_3} and further taking the Taylor expansion on the function $Q^{-1}\left(1-{\epsilon}_{k,\text{co}}^{\text{mrc}}
+{U}^{\text{mrc}}_{k,\text{co}}\right)$ at the point
$1-{\epsilon}_{k,\text{co}}^{\text{mrc}}$, we obtain the achievable coding rate in \eqref{eq_20210305_2}.

Observing \eqref{eq_20210305_2}, it is clear that
the third term decreases with the order $\mathcal{O}\left(\frac{1}{\sqrt{L}}\right)$, while both the second and the last term decrease with the order $\mathcal{O}\left(\frac{1}{L}\right)$. In this context, when the number of blocks is large enough, by ignoring the terms of the order $\mathcal{O}\left(\frac{1}{L}\right)$,  the achievable coding rate $R^{\text{mrc}}_{k,\text{co}}$ in \eqref{eq_20210305_2} is approximated as \eqref{eq_20210923_4} shown at the top of next page,
%
\begin{figure*}[!ht]
\begin{align}\label{eq_20210923_4}
R^{\text{mrc}}_{k,\text{co}}
\approx
R^{\text{mrc},\text{co}}_{k,\text{app}}
=\frac{1}{L}
\sum\limits^{L}_{j=1}I_{k,\text{co}}^{\text{mrc}}(j)
+\frac{\ln\mu}{LT_{c}}
%
-\sqrt{\frac{\frac{1}{L}\sum\limits^{L}_{j=1}
{V}_{k,\text{co}}^{\text{mrc}}(j)}{LT_{c}}}
Q^{-1}\left({\epsilon}_{k,\text{co}}^{\text{mrc}}
\right),
\end{align}
\center\hrulefill
\vspace{-3mm}
\end{figure*}
in which the term $\frac{\ln\mu}{LT_{c}}$ is retained only for the sake of analytical consistency of \eqref{eq_20210923_2} and \eqref{eq_20211014_1} given later in the paper.
This is called the normal approximation and it can accurately characterize the theoretical achievable coding rate for sufficiently large number of blocks. Meanwhile, with \eqref{eq_20210923_4}, the achievable coding rate approaches to the corresponding channel capacity  with a scaling speed of $\frac{1}{\sqrt{LT_{c}}}$, i.e., in the order {\small$\mathcal{O}\left(\frac{1}
{\sqrt{LT_{c}}}\right)$}, and the average error probability can now be approximated as
%
\begin{align}\label{eq_20211014_1}
\epsilon^{\text{mrc}}_{k,\text{co}}
\approx 1-{q}^{\text{mrc},\text{co}}_{k,\text{thres}}
\triangleq \epsilon^{\text{mrc-Taylor}}_{k,\text{co}}.
\end{align}
\end{remark}

Remark \ref{remark_mrc_per} provides clarifications for a varying  number of blocks and fixed number of  BS antennas.
However, it is obvious that the channel capacity $\{I_{k,\text{co}}^{\text{mrc}}(j)\}^{L}_{j=1}$, channel dispersion $\{{V}_{k,\text{co}}^{\text{mrc}}(j)\}^{L}_{j=1}$, and the channel perturbation ${U}^{\text{mrc}}_{k,\text{co}}$ in Proposition \ref{prop_20200103_1} are functions of
$\{{\rm SINR}^{{\text{mrc}}}_{k,\text{co}}(j)\}^{L}_{j=1}$, which from \eqref{sinr_mrc} depend on the fading channels $\{{\bf h}_{k}(j)\}^{L}_{j=1}$, for all $k=1,\ldots,K_{u}$. Since the randomness of the fading channels $\{{\bf h}_{k}(j)\}^{K_{u}}_{k=1}$ (the channel hardening level) strongly relates with the number of the BS antennas,
it is hard to analyze Proposition \ref{prop_20200103_1} in terms of the number of BS antennas directly.
Thus, to see the impact of using large number of BS antennas case and
unveil more system insights, an asymptotic analysis will be performed.
By using the law of large numbers,
the ${\rm SINR}^{{\text{mrc}}}_{k,\text{co}}(j)$ in \eqref{sinr_mrc} will approach to
%
\begin{align}\label{sinr_mrc_large_1}
{\rm SINR}^{{\text{mrc}}}_{k,\text{co}}(j)
&=\frac{P_{k}\frac{1}{N_{b}}\|{\bf h}_{k}(j)\|^{2}_{2}}
{\sum\limits^{K_{u}}_{\substack{m=1\\ m\neq k}}P_{m}
\frac{1}{N_{b}}\left|\frac{{\bf h}^{\dagger}_{k}(j)}{\|{\bf h}_{k}(j)\|_{2}}{\bf h}_{m}(j)\right|^{2}
+\frac{1}{N_{b}}\sigma^{2}_{n}}\notag\\
&\xrightarrow[N_{b} \rightarrow \infty]{a.s.}\infty.
\end{align}
%
By substituting  \eqref{sinr_mrc_large_1} into \eqref{eq_20200103_2}-\eqref{eq_u_1},
the following corollary can be obtained.
\begin{corollary}\label{cor_0824_1}
Without considering any power scaling laws, the asymptotic results (when the number of BS antennas goes to infinity) of the channel capacity $I_{k,\text{co}}^{\text{mrc}}(j)$ in \eqref{eq_20200103_2}, the channel dispersion $V^{\text{mrc}}_{k,\text{co}}(j)$ in \eqref{eq_20200103_3}, and  the channel perturbation term ${U}^{\text{mrc}}_{k,\text{co}}$ in \eqref{eq_u_1} of the achievable coding rate, are:
%
\begin{align}\label{eq_20200810_8}
I_{k,\text{co}}^{\text{mrc}}(j)
\xrightarrow[N_{b} \rightarrow \infty]{a.s.}
\bar{I}_{k,\text{co}}^{\text{mrc}}
=\infty,
\end{align}
%
\begin{align}\label{eq_20200810_8-v}
V^{\text{mrc}}_{k,\text{co}}(j)
\xrightarrow[N_{b} \rightarrow \infty]{a.s.}
\bar{V}_{k,\text{co}}^{\text{mrc}}
=2,
\end{align}
%
\begin{align}\label{eq_0824_1}
{U}^{\text{mrc}}_{k,\text{co}}
&\xrightarrow[N_{b} \rightarrow \infty]{a.s.}
\bar{{U}}^{\text{mrc}}_{k,\text{co}}
\notag\\
&
=\frac{4c_{1}}{\sqrt{2\pi}}
\frac{1}{\sqrt{L}}
\frac{1}{\sqrt{T_{c}}\Gamma\left(T_{c}+1\right)}
\Gamma\left(\frac{2T_{c}+3}{2}\right)\notag\\
&\leq \frac{4c_{1}}{\sqrt{2\pi}}
\frac{1}{\sqrt{L}}
(1+1/T_{c})^{1/2}.
\end{align}
\end{corollary}
\begin{remark}\label{remark_0810_1}
Note that when converting the real case in \cite{collins2018coherent} to the complex case, the asymptotic results of the channel capacity in \eqref{eq_20200810_8} and dispersion in \eqref{eq_20200810_8-v} (without considering the power scaling law) are aligned with the asymptotic result of the channel capacity in  Eq. (273) and channel dispersion in Eq. (274) of \cite{collins2018coherent}. Now,
from Proposition \ref{prop_20200103_1} and Corollary \ref{cor_0824_1}, the following insights can be obtained:
\begin{enumerate}[i)]
\item With massive antenna arrays at the BS, both the channel dispersion $V^{\text{mrc}}_{k,\text{co}}(j)$ and the perturbation  ${U}^{\text{mrc}}_{k,\text{co}}$ of the achievable coding rate approach to a constant. This indicates that for our system setup, in the finite blocklength regime,
    these two important terms cannot be eliminated even when $N_{b}$ goes to infinity.
    This means that the clarifications in Remark \ref{remark_mrc_per} are valid under massive MIMO, i.e., the final effective channel dispersion (the third term in \eqref{eq_20210305_2}) goes to $\sqrt{\frac{2}{LT_{c}}}$, which decreases as the order $\mathcal{O}\left(\frac{1}{\sqrt{L}}\right)$, whilst the final effective perturbation (the last term in \eqref{eq_20210305_2}) still decreases with the order $\mathcal{O}\left(\frac{1}{L}\right)$.
\item For massive MIMO, due to \eqref{eq_20200810_8} and \eqref{eq_20200810_8-v},  the function ${q}^{\text{mrc},\text{co}}_{k,\text{thres}}$ of \eqref{eq_20210923_2} approaches to 1, and the asymptotic result of  the channel perturbation  $\bar{{U}}^{\text{mrc}}_{k,\text{co}}<1$ even with a small number of blocks $L$ and coherence intervals $T_{c}$ (e.g., $L=1$, $T_{c}=5$).  This implies that the condition ${U}^{\text{mrc}}_{k,\text{co}}\leq {q}^{\text{mrc},\text{co}}_{k,\text{thres}}$ can be more easily achieved in massive MIMO systems. Meanwhile, the BE-CLT-based upper bound of the average error probability $\epsilon^{\text{mrc-beclt}}_{k,\text{co}}$ of \eqref{eq_20210917_3} is   approaching to
%
\begin{align}
\epsilon^{\text{mrc-beclt}}_{k,\text{co}}
\xrightarrow[N_{b} \rightarrow \infty]{a.s.}
 \bar{{U}}^{\text{mrc}}_{k,\text{co}},
\end{align}
%
while the Taylor expansion based approximation of the average error probability $\epsilon^{\text{mrc-Taylor}}_{k,\text{co}}$ of \eqref{eq_20211014_1} will converge to zero.  This is a big difference in the massive MIMO regime
while a detailed
comparison between these two results will be presented  in the  numerical results.
\item In terms of \eqref{eq_20210923_4}, as shown in
\cite[Eq. (227)]{polyanskiy2010channel},
the minimum blocklength required for achieving a certain fraction $0<\eta_{k}< 1$ of the channel capacity is as
%
\begin{align}
\label{eq_20200121_6}
n=LT_{c}\approx
\left[\frac{Q^{-1}(\epsilon_{k})}
{1-\eta_{k}}\right]^2
\frac{V_{k}}{\left(C_{k}\right)^2},
\end{align}
%
where we ignore the adjusting term $\frac{\ln\mu}{LT_{c}}$ for convenience.
The terms $C_{k}$ and $V_{k}$ denote the channel capacity and dispersion.
Using  the asymptotic results in Corollary \ref{cor_0824_1}, we infer that the required minimum blocklength can approach to a extremely small value with a large number of BS antennas. This conclusion indicates that massive MIMO is beneficial in reducing the required blocklength, and so is in reducing the communication delay.
\end{enumerate}
\end{remark}

Corollary \ref{cor_0824_1} and Remark \ref{remark_0810_1} are the asymptotic results without considering any power scaling laws.  However, under the power scaling law at the users' side, i.e., the transmit power for each user is scaled with $N_{b}$ according to $P_{k}=\frac{E_{k}}{N_{b}}$, for $k=1,\ldots,K_{u}$, where $E_{k}$ is fixed, the ${\rm SINR}^{{\text{mrc}}}_{k,\text{co}}(j)$ in \eqref{sinr_mrc} will now approach to
%
\begin{align}\label{sinr_mrc_large_power_scaling_co}
{\rm SINR}^{{\text{mrc}}}_{k,\text{co}}(j)
&=\frac{\frac{E_{k}}{N_{b}}\|{\bf h}_{k}(j)\|^{2}_{2}}
{\sum\limits^{K_{u}}_{\substack{m=1\\ m\neq k}}
\frac{E_{m}}{N_{b}}\left|\frac{{\bf h}^{\dagger}_{k}(j)}{\|{\bf h}_{k}(j)\|_{2}}{\bf h}_{m}(j)\right|^{2}
+\sigma^{2}_{n}}\notag\\
&\xrightarrow[N_{b} \rightarrow \infty]{a.s.}
\frac{\gamma^2_{k}E_{k}}{\sigma^{2}_{n}},
\end{align}
%
which is obviously quite different from the asymptotic result given in \eqref{sinr_mrc_large_1}.  Substituting this asymptotic result into \eqref{eq_20200103_2}-\eqref{eq_u_1}
of Proposition~\ref{prop_20200103_1}, we have the following corollary.
\begin{corollary}\label{corollary_co_mrc_power}
Under the power scaling law $P_{k}=\frac{E_{k}}{N_{b}}$, for all $k=1,\ldots,K_{u}$, when $N_{b}\rightarrow \infty$,
the capacity $I_{k,\text{co}}^{\text{mrc}}(j)$, the dispersion $V^{\text{mrc}}_{k,\text{co}}(j)$, and  the perturbation term ${U}^{\text{mrc}}_{k,\text{co}}$ in Proposition \ref{prop_20200103_1}, now approach to
%
\begin{align}\label{eq_20211108_5}
&
I_{k,\text{co}}^{\text{mrc}}(j)
\!\xrightarrow[N_{b} \rightarrow \infty]{a.s.}\!
\bar{\bar{I}}_{k,\text{co}}^{\text{mrc}}
\!=\!\ln\left(1\!+\!\frac{\gamma^2_{k}E_{k}}
{\sigma^{2}_{n}}\right),
\end{align}
%
\begin{align}\label{eq_20211108_5-v}
V^{\text{mrc}}_{k,\text{co}}(j)
\!\xrightarrow[N_{b} \rightarrow \infty]{a.s.}\!
\bar{\bar{V}}_{k,\text{co}}^{\text{mrc}}
\!=\!2\!-\!\frac{2}{1\!+\!\frac{\gamma^2_{k}E_{k}}
{\sigma^{2}_{n}}},
\end{align}
\begin{align}\label{eq_20211108_5-u}
{U}^{\text{mrc}}_{k,\text{co}}
\!\xrightarrow[N_{b} \rightarrow \infty]{a.s.}\!
\bar{{U}}^{\text{mrc}}_{k,\text{co}},
\end{align}
where $\bar{{U}}^{\text{mrc}}_{k,\text{co}}$ has been given in \eqref{eq_0824_1} of Corollary \ref{cor_0824_1}.
\end{corollary}
\begin{remark}
\label{remark_co_mrc_power_scaling}
Under the power scaling law, massive MIMO can suppress the channel dispersion of the achievable coding rate while sacrificing the performance loss of the channel capacity.
 Thus, for the achievable coding rate $R^{\text{mrc},\text{co}}_{k,\text{app}}$ of \eqref{eq_20210923_4} after dropping the terms  $\mathcal{O}\left(\frac{1}{L}\right)$, under the power scaling law, the backoff from the corresponding channel capacity is reduced for fixed average error probability ${\epsilon}_{k,\text{co}}^{\text{mrc}}$ and blocklength $LT_{c}$. Meanwhile, substituting the asymptotic results of Corollary \ref{corollary_co_mrc_power} into \eqref{eq_20210917_3} and \eqref{eq_20211014_1}, we have that both the average error probability $\epsilon^{\text{mrc-beclt}}_{k,\text{co}}$ in \eqref{eq_20210917_3} and $\epsilon^{\text{mrc-Taylor}}_{k,\text{co}}$ in \eqref{eq_20211014_1} cannot go to zero even with massive MIMO since the function ${q}^{\text{mrc},\text{co}}_{k,\text{thres}}$ in \eqref{eq_20210923_2} and ${U}^{\text{mrc}}_{k,\text{co}}$ are approaching to constants. Also, the minimum required blocklength in \eqref{eq_20200121_6}
 will also approach to a constant. Therefore, the conclusions are quite different  with and without considering the power scaling laws.

 Furthermore, from Proposition \ref{prop_20200103_1} and Corollaries \ref{cor_0824_1} \& \ref{corollary_co_mrc_power}, we can infer that fixing the total blocklength $n=LT_{c}$, both the analytical result ${U}^{\text{mrc}}_{k,\text{co}}$ and the asymptotic result $\bar{{U}}^{\text{mrc}}_{k,\text{co}}$ of the channel perturbation term increase along with the coherence interval $T_{c}$.  However, from  Proposition \ref{prop_20200103_1}, leveraging more number of blocks can not only help to reduce the channel perturbation, but also help to reduce the performance uncertainty caused by the randomness of the channels. In this sense, we can conclude that in the coherent case, the scenarios with higher number of blocks and  shorter coherence intervals are more beneficial.
\end{remark}
\vspace{-3mm}
\subsection{ZF Scheme}\label{zf_transmission}
With ZF scheme, we must have $K_u \leq N_b$. Let ${\bf H}(j)=[{\bf h}_{1}(j),\ldots,{\bf h}_{K_{u}}(j)]\in\mathcal{C}^{N_{b}\times K_{u}}$ be the channel matrix from all users to the BS, and ${\bf X}(j)=[{\bf x}_{1}(j),\ldots,{\bf x}_{K_{u}}(j)]\in \mathcal{C}^{T_{c}\times K_{u}}$ be the input signal matrix from all users. Then, the ZF vector ${\bf a}_{k}(j)={\bf a}_{k,\text{co}}^{\text{zf}}(j)$, is given by
%
\begin{align}\label{eq_20200106_1}
{\bf a}_{k,\text{co}}^{\text{zf}}(j)
&=\frac{\left({\bf H}(j)({\bf H}^{\dagger}(j){\bf H}(j))^{-1}\right)_{:,k}}{\|({\bf H}(j)\left({\bf H}^{\dagger}(j){\bf H}(j)\right)^{-1})_{:,k}\|_{2}}\notag\\
&=\frac{\left({\bf h}^{\dagger}_{k}(j){\bf V}_{k}(j)\right)^{\dagger}}{\|{\bf h}^{\dagger}_{k}(j){\bf V}_{k}(j)\|_{2}},
\end{align}
%
where ${\bf V}_{k}(j)$ is the null projection of the matrix ${\bf H}_{\slashed{k}}(j)$, which is given by
%
\begin{align}\label{eq_zf_20210312_2}
{\bf V}_{k}(j)={\bf I}_{N_{b}}-{\bf H}_{\slashed{k}}(j)\left({\bf H}^{\dagger}_{\slashed{k}}(j){\bf H}_{\slashed{k}}(j)\right)^{-1}{\bf H}^{\dagger}_{\slashed{k}}(j),
\end{align}
%
in which ${\bf H}_{\slashed{k}}(j)$ is the submatrix obtained by removing ${\bf h}_{k}(j)$ from ${\bf H}(j)$.
Substituting the  ZF vector  of \eqref{eq_20200106_1} into \eqref{eq_20191015_1} and taking into account
${\bf V}_{k}(j){\bf H}_{\slashed{k}}(j)={\bf 0}_{N_{b}}$, we obtain
%
\begin{align}\label{eq_20191031_4}
\tilde{{\bf y}}^{\text{zf}}_{k,\text{co}}(j)
&=\underbrace{\|{\bf h}^{\dagger}_{k}(j){\bf V}_{k}(j)\|_{2}{\bf x}^{T}_{k}(j)}_{\text{Desired signal}}\notag\\
&\quad+\underbrace{\frac{{\bf h}^{\dagger}_{k}(j){\bf V}_{k}(j)}{\|{\bf h}^{\dagger}_{k}(j){\bf V}_{k}(j)\|_{2}}{\bf N}(j)}_{\text{Colored noise}}.
\end{align}
%

We denote the noise vector in \eqref{eq_20191031_4} as $\tilde{{\bf n}}_{\text{co}}^{\text{zf}}(j)=\frac{{\bf h}^{\dagger}_{k}(j){\bf V}_{k}(j)}{\|{\bf h}^{\dagger}_{k}(j){\bf V}_{k}(j)\|_{2}}{\bf N}(j)\in \mathcal{C}^{1\times T_{c}}$.
Thus, for the coherent case,  \eqref{eq_20191031_4} can also be treated as a point-to-point SISO channel with the fading channel $\|{\bf h}^{\dagger}_{k}(j){\bf V}_{k}(j)\|_{2}$ and independent Gaussian noise $\tilde{{\bf n}}_\text{co}^{\text{zf}}(j)\sim \mathcal{CN}({\bf0},\sigma^{2}_{n}{\bf I}_{T_{c}})$. The corresponding signal-to-noise ratio with the ZF scheme is
%
\begin{align}\label{zf_eq_sinr}
{\rm SNR}^{\text{zf}}_{k,\text{co}}(j)
=\frac{P_{k}}{
\sigma^{2}_{n}}\|{\bf h}^{\dagger}_{k}(j){\bf V}_{k}(j)\|^{2}_{2}.
\end{align}

So, by replacing the terms $\|{\bf h}_{k}(j)\|_{2}$ and $\sum\limits^{K_{u}}_{\substack{m=1\\ m\neq k}}P_{m}\left|\frac{{\bf h}^{\dagger}_{k}(j)}{\|{\bf h}_{k}(j)\|_{2}}{\bf h}_{m}(j)\right|^{2}+\sigma^{2}_{n}$ of the MRC scheme in Proposition~\ref{prop_20200103_1}  of Section \ref{mrc_scheme_trans} with $\|{\bf h}^{\dagger}_{k}(j){\bf V}_{k}(j)\|_{2}$ and $\sigma^{2}_{n}$, respectively,
the following proposition  for the ZF scheme under the finite blocklength regime can be obtained.
\begin{prop}
\label{cor_20200122_1}
Under the assumptions of coherent block-fading channels and the i.i.d. Gaussian input signals,
with the ML decoder at the BS,
 define a threshold term  as
%
\begin{align}\label{eq_20211106_1}
{q}^{\text{zf},\text{co}}_{k,\text{thres}}
&\triangleq
Q\left(\left(\frac{\frac{1}{L}\sum\limits^{L}_{j=1}
V_{k,\text{co}}^{\text{zf}}(j)}
{LT_{c}}\right)^{-\frac{1}{2}}\right.\notag\\
&\left.\quad\quad\times\left(\frac{\ln M_{k}}{LT_{c}}-\frac{\ln\mu}{LT_{c}}
-\frac{1}{L}
\sum\limits^{L}_{j=1}
I_{k,\text{co}}^{\text{zf},\text{im}}(j)\right)\right).
\end{align}
%
When the channel perturbation term $U^{\text{zf}}_{k,\text{co}}$, which will be given later in \eqref{eq_zf_0811_2}, satisfies the condition $U^{\text{zf}}_{k,\text{co}}\leq {q}^{\text{zf},\text{co}}_{k,\text{thres}}$, the relation between the achievable coding rate $R^{\text{zf},\text{co}}_{k}$ and the average error probability ${\epsilon}_{k,\text{co}}^{\text{zf}}$ is given by:
%
\begin{align}
\label{eq_20200106_4}
&\quad\frac{\ln M_{k}}{LT_{c}}
\geq
R^{\text{zf},\text{co}}_{k}
\!\triangleq\!
\frac{1}{L}
\sum\limits^{L}_{j=1}I_{k,\text{co}}^{\text{zf}}(j)
+\frac{\ln\mu}{LT_{c}}\notag\\
&-\sqrt{\frac{\frac{1}{L}\sum\limits^{L}_{j=1}
{V}_{k,\text{co}}^{\text{zf}}(j)}{LT_{c}}}
Q^{-1}\left({\epsilon}_{k,\text{co}}^{\text{zf}}
\right)+\sqrt{\frac{\frac{1}{L}\sum\limits^{L}_{j=1}
{V}_{k,\text{co}}^{\text{zf}}(j)}{LT_{c}}}
\mathcal{O}\left({U}^{\text{zf}}_{k,\text{co}}
\right).
\end{align}
%
In \eqref{eq_20211106_1} and \eqref{eq_20200106_4}, the parameter $\mu$ is also uniformly distributed on the interval $(0,1)$.
The terms $I_{k,\text{co}}^{\text{zf}}(j)$ and  $V_{k,\text{co}}^{\text{zf}}(j)$ denote the channel capacity and dispersion at the $j$-th block, which are given by
%
\begin{align}\label{eq_20200106_5}
I_{k,\text{co}}^{\text{zf}}(j)
=\ln\left(1+\frac{P_{k}}{
\sigma^{2}_{n}}\|{\bf h}^{\dagger}_{k}(j){\bf V}_{k}(j)\|^{2}_{2}\right),
\end{align}
\begin{align}\label{eq_20200106_6}
V^{\text{zf}}_{k,\text{co}}(j)
=2\left(1-\frac{1}
{1+\frac{P_{k}}{\sigma^{2}_{n}}\|{\bf h}^{\dagger}_{k}(j){\bf V}_{k}(j)\|^{2}_{2}}\right).
\end{align}
%
The term ${U}^{\text{zf}}_{k,\text{co}}$ is  the channel perturbation and has the form
%
\begin{align}\label{eq_zf_0811_2}
U^{\text{zf}}_{k,\text{co}}
&=\frac{4c_{1}}{\sqrt{2\pi}}
\frac{1}{\sqrt{LT_{c}}\Gamma\left(T_{c}+1\right)}
\Gamma\left(\frac{2T_{c}+3}{2}\right)\notag\\
&\quad\times\left(\!\frac{1}{L}
\sum\limits^{L}_{j=1}{V}_{k,\text{co}}^{\text{zf}}(j)
\!\right)^{-\frac{3}{2}}\!\!
\left\{\frac{1}{L}\sum\limits^{L}_{j=1}
\left[V^{\text{zf}}_{k,\text{co}}(j)
\right]^{\frac{3}{2}}\right\},
\end{align}
\end{prop}
%
\begin{remark}\label{remark_co_zf_1}
Observing Proposition \ref{cor_20200122_1}
and Proposition \ref{prop_20200103_1}, the only difference between the ZF and MRC scheme for the coherent case is the {\rm SINR} in \eqref{zf_eq_sinr} and \eqref{sinr_mrc}. Hence, all the conclusions obtained in Remark \ref{remark_mrc_per} also hold for the ZF scheme. Likewise, we have:
\begin{enumerate}[i)]
\item The exact difference between the finite blocklength and infinite blocklength comes from the channel dispersion, channel perturbation, and the non-zero average decoding error probability. From Propositions \ref{prop_20200103_1} \& \ref{cor_20200122_1}, these terms, which do not contribute to the improvement of the achievable coding rate, can be made arbitrarily small by letting
the number of blocks tend to infinity.
Meanwhile, similarly as in the MRC scheme, the achievable coding rate for the ZF scheme after ignoring the terms of the order $\mathcal{O}\left(\frac{1}{L}\right)$,
is given by
%
\begin{align}\label{eq_20211108_1}
R^{\text{zf}}_{k,\text{co}}
\approx
R^{\text{zf},\text{co}}_{k,\text{app}}
&=
\frac{1}{L}
\sum\limits^{L}_{j=1}I_{k,\text{co}}^{\text{zf}}(j)
+\frac{\ln\mu}{LT_{c}}\notag\\
&\quad-\sqrt{\frac{\frac{1}{L}\sum\limits^{L}_{j=1}
{V}_{k,\text{co}}^{\text{zf}}(j)}{LT_{c}}}
Q^{-1}\left({\epsilon}_{k,\text{co}}^{\text{zf}}
\right),
\end{align}
%
and  the average error probability  $\epsilon^{\text{zf}}_{k,\text{co}}$ after using the BE-CLT and the Taylor approximation are respectively as 
%
\begin{align}\label{eq_20211108_2}
&\epsilon^{\text{zf}}_{k,\text{co}}
\leq 1-{q}^{\text{zf},\text{co}}_{k,\text{thres}}
+{U}^{\text{zf}}_{k,\text{co}}
\triangleq \epsilon^{\text{zf-beclt}}_{k,\text{co}},
\end{align}
%
\begin{align}\label{eq_20211108_2-v}
\epsilon^{\text{zf}}_{k,\text{co}}
\approx 1-{q}^{\text{zf},\text{co}}_{k,\text{thres}}
\triangleq \epsilon^{\text{zf-Taylor}}_{k,\text{co}}.
\end{align}
From \eqref{eq_20211108_1} \& \eqref{eq_20211108_2} \& \eqref{eq_20211108_2-v} for the ZF scheme and \eqref{eq_20210917_3} \& \eqref{eq_20210923_4} \&
\eqref{eq_20211014_1} for the MRC scheme,
we can observe that the performance of both schemes is determined
 by the interaction between the average capacity,   average dispersion,  average perturbation through all blocks, and the total blocklength ($L$ and $T_{c}$). Thus, it is challenging, if not impossible, to conjecture which scheme performs better in the finite blocklength space. A detailed numerical comparison is given in the numerical results section;
\item Similar to Corollaries \ref{cor_0824_1} \& \ref{corollary_co_mrc_power}, the asymptotic analysis can also be performed with massive MIMO  by using the following asymptotic result,
%
\begin{align}
\label{eq_20210306}
\frac{1}{N_{b}}\|{\bf h}^{\dagger}_{k}(j){\bf V}_{k}(j)\|^{2}_{2}
\!\xrightarrow[N_{b}\rightarrow \infty]{a.s.}\!
\gamma^{2}_{k},
\end{align}
%
from which, it is easy to see that the power loss in the ZF scheme can be eliminated with massive MIMO.
Now, by substituting \eqref{eq_20210306} into \eqref{zf_eq_sinr} and further into \eqref{eq_20200106_5}-\eqref{eq_zf_0811_2},
the same asymptotic results as that given in  Corollary \ref{cor_0824_1} without any power scaling law and Corollary \ref{corollary_co_mrc_power}  under the power scaling law $P_{k}=\frac{E_{k}}{N_{b}}$, for all $k=1,\ldots,K_{u}$,  can also be obtained for the ZF scheme.  Thus, all the conclusions obtained in Remarks \ref{remark_0810_1} \& \ref{remark_co_mrc_power_scaling}
also hold  here;
\item The instantaneous channel capacity, channel dispersion, and channel perturbation terms during a certain block $j$, are expressed as the realizations of the uplink channels in that block, and the corresponding three key terms in the achievable coding rate under block fading channels are the average values across the whole number of blocks.
\end{enumerate}
\end{remark}
\vspace{-4mm}
\section{Non-coherent Channels}
\label{section_imperfect}
In this section, non-coherent channels are considered, i.e., the BS has no perfect CSI of the uplink channels and it needs to estimate them firstly with the received pilots at the beginning of each coherent block. We consider the usual time-division duplex (TDD) scheme and the MMSE channel estimation is applied at the BS. Thus, each coherence interval $T_{c}$ is divided into two phases:
the channel estimation phase, in which  $\tau_{c}$ symbols are used for the transmission of orthogonal pilot sequences simultaneously by all users, and the signal transmission phase with the remaining $T_{c}-\tau_{c}$ symbols in each block being used for signals transmission. Assume that the $k$-th user transmits the pilot signal with the transmit power $P^{c}_{k}$.
Then, the uplink channel in  $j$-th block can be written as
%
\begin{align}\label{eq_im_20210306_1}
{\bf h}_{k}(j)
=\hat{{\bf h}}_{k}(j)+\tilde{{\bf h}}_{k}(j),
\end{align}
%
for all $k=1,\ldots,K_{u}$ and $j=1,\ldots,L$, In this equation,
$\hat{{\bf h}}_{k}(j)$ and $\tilde{{\bf h}}_{k}(j)$ are independent and denote the estimated channel and the estimation error at the BS with the distribution $\hat{{\bf h}}_{k}(j)\sim \mathcal{CN}\left(0, \phi_{k}^2{\bf I}_{N_{b}}\right)$  and $\tilde{{\bf h}}_{k}(j)\sim \mathcal{CN}\left(0, (\gamma^{2}_{k}-\phi_{k}^2){\bf I}_{N_{b}}\right)$, respectively. The coefficient $\phi_{k}^2$, representing the estimated channel gain,
is given by
%
\begin{align}\label{eq_im_20210306_2}
\phi_{k}^2
=\gamma^{4}_{k}\left(\gamma^{2}_{k}
+\frac{\sigma^{2}_{c}}{\tau_{c}P^{c}_{k}}
\right)^{-1},
\end{align}
%
where $k=1,\ldots,K_{u}$. The notation
$\sigma^{2}_{c}$ denotes the variance of the AWGN at the channel estimation phase.
\vspace{-4mm}
\subsection{MRC Scheme}\label{section_im_mrc}
For the non-coherent case,
 the vector ${\bf a}_{k}(j)$ at the $j$-th block in \eqref{eq_20191015_1} for the MRC scheme is
%
\begin{align}\label{eq_2021_0211_1}
{\bf a}_{k,\text{non}}^{\text{mrc}}(j)
=\frac{\hat{{\bf h}}_{k}(j)}{\|\hat{{\bf h}}_{k}(j)\|_{2}},
\end{align}
%
for all $k=1,\ldots,K_{u}$ and $j=1,\ldots,L$.
Substituting \eqref{eq_2021_0211_1} and the estimated channel of \eqref{eq_im_20210306_1} into \eqref{eq_20191015_1} yields
%
\begin{align}\label{eq_im_20210307_2}
\tilde{{\bf y}}^{\text{mrc}}_{k,\text{non}}(j)
=\|\hat{{\bf h}}^{\dagger}_{k}(j)\|_{2}{\bf x}^{T}_{k}(j)
+\check{{\bf n}}^{\text{mrc}}_{\text{non}}(j),
\end{align}
%
where $\tilde{{\bf y}}^{\text{mrc}}_{k,\text{non}}(j)\in \mathcal{C}^{1\times(T_c-\tau_{c})}$ and
the input signals ${\bf x}_{k}(j)\in \mathcal{C}^{(T_c-\tau_{c})\times 1}$ is distributed as  ${\bf x}_{k}(j)\sim \mathcal{CN}\left({\bf 0}, P_{k}{\bf I}_{T_{c}-\tau_{c}}\right)$. The term $\check{{\bf n}}^{\text{mrc}}_{\text{non}}(j)$ denotes the effective noise, which has the form
%
\begin{align}\label{eq_im_20210307_1}
\check{{\bf n}}^{\text{mrc}}_{\text{non}}(j)
&\!\triangleq\!
\frac{\hat{{\bf h}}^{\dagger}_{k}(j)}{\|\hat{{\bf h}}^{\dagger}_{k}(j)\|_{2}}\tilde{{\bf h}}_{k}(j){\bf x}^{T}_{k}(j)
%
\!+\!\sum\limits^{K_{u}}_{\substack{m=1\\ m\neq k}}\frac{\hat{{\bf h}}^{\dagger}_{k}(j)}{\|\hat{{\bf h}}^{\dagger}_{k}(j)\|_{2}}\hat{{\bf h}}_{m}(j){\bf x}^{T}_{m}(j)\notag\\
&\quad\!+\!\sum\limits^{K_{u}}_{\substack{m=1\\ m\neq k}}\frac{\hat{{\bf h}}^{\dagger}_{k}(j)}{\|\hat{{\bf h}}^{\dagger}_{k}(j)\|_{2}}\tilde{{\bf h}}_{m}(j){\bf x}^{T}_{m}(j)
\!+\!\frac{\hat{{\bf h}}^{\dagger}_{k}(j)}{\|\hat{{\bf h}}^{\dagger}_{k}(j)\|_{2}}{\bf N}(j).
\end{align}
%
After the channel estimation phase, the BS treats the estimated channels $\{\hat{{\bf h}}_{k}(j)\}^{K_{u}}_{k=1}$ as the true channels. However, the channel estimation errors $\{\tilde{{\bf h}}_{k}(j)\}^{K_{u}}_{k=1}$ are unknown at the BS, which leads to the fact that the first and third terms in \eqref{eq_im_20210307_1} are not Gaussian distributed
vectors. Thus, the  effective noise $\check{{\bf n}}^{\text{mrc}}_{\text{non}}(j)\in \mathcal{C}^{1\times (T_{c}-\tau_{c})}$ in \eqref{eq_im_20210307_2} is also not Gaussian distributed, which has zero mean and  covariance matrix ${\bf \Sigma}_{\check{{\bf n}}^{\text{mrc}}_{\text{non}}}(j)
=\sigma^{2}_{\check{{\bf n}}^{\text{mrc}}_{\text{non}}}(j){\bf I}_{T_{c}-\tau_{c}}$,
in which $\sigma^{2}_{\check{{\bf n}}^{\text{mrc}}_{\text{non}}}(j)$ is
%
\begin{align}\label{eq_im_20210307_4}
\sigma^{2}_{\check{{\bf n}}^{\text{mrc}}_{\text{non}}}(j)
&=\sum\limits^{K_{u}}_{k=1}P_{k}(\gamma^{2}_{k}-\phi_{k}^2)
\notag\\
&\quad+\sum\limits^{K_{u}}_{\substack{m=1\\ m\neq k}}P_{m}\left[\frac{\left|\hat{{\bf h}}^{\dagger}_{m}(j)\hat{{\bf h}}_{k}(j)\right|^{2}}{\|\hat{{\bf h}}^{\dagger}_{k}(j)\|^{2}_{2}}\right]
+\sigma^{2}_{n}.
\end{align}
%
This is quite different from the corresponding coherent case, in which the effective noise is Gaussian distributed.  Hence, the non-Gaussian effective noise makes the non-coherent case more challenging and difficult to analyze.

Now, \eqref{eq_im_20210307_2} can be treated as a point-to-point SISO channel with the coherent fading channel $\|\hat{{\bf h}}^{\dagger}_{k}(j)\|_{2}$, the i.i.d. Gaussian input signal vector ${\bf x}^{T}_{k}(j)$ and the non-Gaussian noise vector $\check{{\bf n}}^{\text{mrc}}_{\text{non}}(j)$. The corresponding signal-to-interference-plus-noise-ratio ${\rm SINR}_{k,\text{non}}^{{\text{mrc}}}(j)$ at the $j$-th block is
\begin{align}\label{sinr_mrc_im}
{\rm SINR}_{k,\text{non}}^{{\text{mrc}}}(j)
=\frac{P_{k}\|\hat{{\bf h}}^{\dagger}_{k}(j)\|^{2}_{2}}{\sigma^{2}_{\check{{\bf n}}^{\text{mrc}}_{\text{non}}}(j)},
\end{align}
%
which reveals the impact of the inter-user interference  and the channel estimation errors.

As shown in Appendix \ref{appendix_prop_imper_mrc}, to determine the achievable coding rate under the non-coherent case, we need to identify the distributions of the normalized squared norms of the output signal vector $\tilde{{\bf y}}^{\text{mrc}}_{k,\text{non}}(j)$ and the effective noise vector $\check{{\bf n}}^{\text{mrc}}_{\text{non}}(j)$ of \eqref{eq_im_20210307_2}. Since it is really hard to derive the exact probability density function (PDF) of these two parameters, we consider their approximations, which are widely used in the literature and shown to be very tight in the numerical results. More precisely, let us define the random variables:
%
\begin{align}
&v^{\text{mrc}}_{y,\text{non}}(j)
\triangleq\frac{\|\tilde{{\bf y}}^{\text{mrc}}_{k,\text{non}}(j)\|_{2}^{2}}
{P_{k}\|\hat{{\bf h}}^{\dagger}_{k}(j)\|^{2}_{2}+\sigma^{2}_{\check{{\bf n}}^{\text{mrc}}_{\text{non}}}(j)},\label{eq_v_yn_mrc}\\
\quad
&v^{\text{mrc}}_{n,\text{non}}(j)
\triangleq\frac{\|\check{{\bf n}}^{\text{mrc}}_{\text{non}}(j)\|^{2}_{2}}{\sigma^{2}_{\check{{\bf n}}^{\text{mrc}}_{\text{non}}}(j)},\label{eq_v_yn_mrc-n}
\end{align}
%
%
which represent the normalized squared norms of the output signal vector and the  effective noise vector, respectively.  Conditioned on $\{\hat{{\bf h}}_{k}(j)\}^{K_{u}}_{k=1}$,
the distribution of $v^{\text{mrc}}_{y,\text{non}}(j)$ and $v^{\text{mrc}}_{n,\text{non}}(j)$ can be approximated by a Gamma distribution by matching the first and second-order moments, which are as follows:\footnote{The reasons for using the Gamma distribution are that: i) it is a Type-III Pearson distribution which
is widely used in approximating distributions of positive random variables by matching the high-order moments \cite{al2010approximation}; ii) its bivariate PDF is also simple and does not involve any higher-order complicated mathematical functions \cite{Shakir2013}; iii) the positive parameters $v^{\text{mrc}}_{\text{y, non}}(j)$ and $v^{\text{mrc}}_{\text{n, non}}(j)$ can be expressed as a summation of a series of squared-norm random variables which are chi-squared like form, while the Gamma distribution is a generalization of the chi-square distribution.}
%
\begin{align}
&v^{\text{mrc}}_{y,\text{non}}(j)\sim \Gamma\left(\beta^{{\text{mrc}}}_{y,\text{non}}(j),
\frac{\beta^{{\text{mrc}}}_{y,\text{non}}(j)}
{T_{c}-\tau_{c}}\right),\label{eq_v_yn_gamma}\\
&v^{\text{mrc}}_{n,\text{non}}(j)\sim \Gamma\left(\beta^{{\text{mrc}}}_{n,\text{non}}(j),
\frac{\beta^{{\text{mrc}}}_{n,\text{non}}(j)}
{T_{c}-\tau_{c}}\right),\label{eq_v_yn_gamma-n}
\end{align}
with the correlation coefficient $\rho_{\text{non}}^{\text{mrc}}(j)$ between $v^{\text{mrc}}_{y,\text{non}}(j)$ and $v^{\text{mrc}}_{n,\text{non}}(j)$ as
%
\begin{align}\label{eq_20210930_1}
\rho_{\text{non}}^{\text{mrc}}(j)
=\frac{\sqrt{\beta^{{\text{mrc}}}_{y,\text{non}}(j)
\beta^{{\text{mrc}}}_{n,\text{non}}(j)}}
{\Xi_{k,\text{non}}^{{\text{mrc}}}(j)},
\end{align}
%
where $\Xi_{k,\text{non}}^{{\text{mrc}}}(j)$ and $\beta^{{\text{mrc}}}_{y,\text{non}}(j)$ are respectively given in \eqref{eq_im_mrc_20210312_1} and \eqref{eq_im_mrc_20210309_20} at the top of this page. The term $\beta^{{\text{mrc}}}_{n,\text{non}}(j)$ is as
%
\begin{figure*}[!ht]
\begin{align}\label{eq_im_mrc_20210312_1}
\Xi_{k,\text{non}}^{{\text{mrc}}}(j)
=\frac{(T_{c}-\tau_{c})
\left({\rm SINR}_{k,\text{non}}^{{\text{mrc}}}(j)+1\right)}
{\left[\left({\rm SINR}_{k,\text{non}}^{{\text{mrc}}}(j)\right)
\left(\frac{P_{k}(\gamma^{2}_{k}-\phi^{2}_{k})}
{\sigma^{2}_{\check{{\bf n}}_{\text{non}}^{\text{mrc}}}(j)}\right)
+1\right]
+\frac{\Delta^{\text{mrc}}_{k,\text{non}}(j)}
{\sigma^{4}_{\check{{\bf n}}_{\text{non}}^{\text{mrc}}}(j)}
\left[(T_{c}-\tau_{c})+1\right]},
\end{align}
\center\hrulefill
\vspace{-3mm}
\end{figure*}
%
\begin{figure*}[!ht]
\begin{align}\label{eq_im_mrc_20210309_20}
\beta^{{\text{mrc}}}_{y,\text{non}}(j)
=\frac{\left(1+{\rm SINR}_{k,\text{non}}^{{\text{mrc}}}(j)\right)^{2}(T_{c}-\tau_{c})}
{\left(1+{\rm SINR}_{k,\text{non}}^{{\text{mrc}}}(j)\right)^{2}
+\left[2\left({\rm SINR}_{k,\text{non}}^{{\text{mrc}}}(j)\right)
\left(\frac{P_{k}(\gamma^{2}_{k}-\phi^{2}_{k})}
{\sigma^{2}_{\check{{\bf n}}_{\text{non}}^{\text{mrc}}}(j)}\right)
+\frac{\Delta^{\text{mrc}}_{k,\text{non}}(j)}{\sigma^{4}_{\check{{\bf n}}_{\text{non}}^{\text{mrc}}}(j)}\right]
\left[(T_{c}-\tau_{c})+1\right]}.
\end{align}
\center\hrulefill
\vspace{-3mm}
\end{figure*}
%
\begin{align}\label{eq_im_mrc_20210312_2}
\beta^{{\text{mrc}}}_{n,\text{non}}(j)
=\frac{(T_{c}-\tau_{c})}
{1+\frac{\Delta^{\text{mrc}}_{k,\text{non}}(j)}
{\sigma^{4}_{\check{{\bf n}}_{\text{non}}^{\text{mrc}}}(j)}
\left[(T_{c}-\tau_{c})+1\right]},
\end{align}
%
In \eqref{eq_im_mrc_20210312_1}, \eqref{eq_im_mrc_20210309_20}, and \eqref{eq_im_mrc_20210312_2},
the term $\Delta^{\text{mrc}}_{k,\text{non}}(j)$ is as
\begin{align}\label{eq_20210311_6}
\Delta^{\text{mrc}}_{k,\text{non}}(j)
&=\sum\limits^{K_{u}}_{k=1}
P^{2}_{k}(\gamma^{2}_{k}-\phi^{2}_{k})^{2}\notag\\
&\quad+2\sum\limits^{K_{u}}_{\substack{m=1\\ m\neq k}}
P^{2}_{m}(\gamma^{2}_{m}-\phi^{2}_{m})
\frac{|\hat{{\bf h}}^{\dagger}_{k}(j)\hat{{\bf h}}_{m}(j)|^{2}}{\|\hat{{\bf h}}^{\dagger}_{k}(j)\|^{2}_{2}}.
\end{align}

Now, under this case, the mismatched nearest-neighbor (NN) decoding rule is applied at the BS.\footnote{Note that when the effective noise is non-Gaussian distributed, it is much more complicated to use the ML decoding metric again since it is quite involved and difficult to calculate the corresponding PDF  as in \eqref{eq_20200731_1}. Hence, 
it is more convenient to deploy the mismatched NN decoding metric, as shown in \cite{scarlett2020information}.}  With \cite{ostman2021urllc}, for arbitrary $s>0$, the generalized mismatched NN decoding rule is given as \eqref{eq_p_zf_20210307_1} shown at the top of this page,
%
%
\begin{figure*}[!ht]
\begin{align}\label{eq_p_zf_20210307_1}
\left(\hat{{\bf X}}_{k}\right)^{L}
&=\underset{\left({\bf X}_{k}\right)^{L}\in\mathcal{C}^{T_{c}\times L}}
{\arg \max} \exp\left\{-s\left\|((\tilde{{\bf Y}}_{k})_{\text{non}}^{\text{mrc}})^{L}-(\hat{{\bf H}}_{k})^{L}(({\bf X}_{k})^{L})^{T}\right\|^{2}_{F}\right\}\notag\\
&=\underset{\left({\bf X}_{k}\right)^{L}\in\mathcal{C}^{T_{c}\times L}}
{\arg \max}
\prod\limits^{L}_{j=1}\exp\left\{-s
\left\|\tilde{{\bf y}}^{\text{mrc}}_{k,\text{non}}(j)-\|\hat{{\bf h}}^{\dagger}_{k}(j)\|_{2}{\bf x}^{T}_{k}(j)\right\|^{2}_{2}
\right\},
\end{align}
\center\hrulefill
\vspace{-2mm}
\end{figure*}
where $(\hat{{\bf H}}_{k})^{L}=\operatorname{diag}\left\{\|\hat{{\bf h}}^{\dagger}_{k}(1)\|_{2},\ldots,\|\hat{{\bf h}}^{\dagger}_{k}(L)\|_{2}\right\}$, for all $k=1,\ldots, K_{u}$, which is perfectly known at the BS. The matrices $({\bf X}_{k})^{L}\in\mathcal{C}^{(T_{c}-\tau_{c})\times L}$  and $\left((\tilde{{\bf Y}}_{k})_{\text{non}}^{\text{mrc}}\right)^{L}
\in\mathcal{C}^{L\times (T_{c}-\tau_{c})}$ have similar definitions as in Section \ref{section_2}.
From Appendix \ref{appendix_prop_imper_mrc},
when choosing $s=\frac{1}{\sigma^{2}_{\check{{\bf n}}^{\text{mrc}}_{\text{non}}}(j)}$ in \eqref{eq_p_zf_20210307_1},  the capacity can be achieved if we use this decoding rule.\footnote{Generally, the i.i.d. Gaussian input signals are not capacity-achieving in mismatched decoding schemes.} Recalling the matched ML decoding metric in \eqref{eq_p_mrc_20210307_1} under the coherent case, the mismatched NN decoding metric in \eqref{eq_p_zf_20210307_1} is quite different from \eqref{eq_p_mrc_20210307_1}, which is exactly due to that the  effective noise in the non-coherent case is not Gaussian distributed. %
Thus, with this decoding metric and setting $s=\frac{1}{\sigma^{2}_{\check{{\bf n}}^{\text{mrc}}_{\text{non}}}(j)}$, the following Proposition \ref{prop_imper_mrc} can be obtained. Before presenting this proposition, some coefficients are first defined as follows:
%
\begin{align}
\beta^{{\text{mrc}}}_{\max,\text{non}}(j)
\triangleq \max\left\{\beta^{{\text{mrc}}}_{y,\text{non}}(j)
,\beta^{{\text{mrc}}}_{n,\text{non}}(j)
\right\},
\end{align}
%
\begin{align}
\beta^{{\text{mrc}}}_{\min,\text{non}}(j)
\triangleq\min\left\{\beta^{{\text{mrc}}}_{y,\text{non}}(j)
,\beta^{{\text{mrc}}}_{n,\text{non}}(j)
\right\},
\end{align}
%
\begin{align}\label{eq_20211028_1}
l^{\text{mrc}}_{11,\text{non}}(j)
&\triangleq \frac{T_{c}-\tau_{c}}
{\sqrt{\beta^{{\text{mrc}}}_{\min,\text{non}}(j)}},
%
%
\end{align}
%
\begin{align}\label{eq_20211028_1-12}
l^{\text{mrc}}_{12,\text{non}}(j)
\triangleq \frac{\left(\sqrt{\beta^{{\text{mrc}}}_{\max,\text{non}}(j)}
-\rho_{\text{non}}^{\text{mrc}}(j)
\sqrt{\beta^{{\text{mrc}}}_{\min,\text{non}}(j)}
\right)^2}{1-(\rho_{\text{non}}^{\text{mrc}}(j))^2},
\end{align}
\begin{align}\label{eq_20211028_1-21}
l^{\text{mrc}}_{21,\text{non}}(j)
&\triangleq \rho_{\text{non}}^{\text{mrc}}(j)
\frac{T_{c}-\tau_{c}}
{\sqrt{\beta^{{\text{mrc}}}_{\max,\text{non}}(j)}},
\end{align}
\begin{align}\label{eq_20211028_1-22}
l^{\text{mrc}}_{22,\text{non}}(j)
\triangleq \sqrt{1-\left(\rho_{\text{non}}^{\text{mrc}}(j)\right)^2}
\frac{T_{c}-\tau_{c}}
{\sqrt{\beta^{{\text{mrc}}}_{\max,\text{non}}(j)}},
\end{align}
\begin{align}\label{eq_20211028_1-total}
l^{\text{mrc}}_{\text{non}}(j)
&\triangleq l^{\text{mrc}}_{22,\text{non}}(j)
\sqrt{\beta^{{\text{mrc}}}_{\min,\text{non}}(j)}\notag\\
&\quad+\left(l^{\text{mrc}}_{11,\text{non}}(j)
-l^{\text{mrc}}_{21,\text{non}}(j)
\right)\sqrt{l^{\text{mrc}}_{12,\text{non}}(j)}.
\end{align}
%
\begin{prop}\label{prop_imper_mrc}
For non-coherent fading channels and i.i.d. Gaussian input signals, when the mismatched NN decoding metric in \eqref{eq_p_zf_20210307_1} is applied at the BS, for the MRC scheme, setting $s=\frac{1}{\sigma^{2}_{\check{{\bf n}}^{\text{mrc}}_{\text{non}}}(j)}$, we define the threshold value ${q}^{\text{mrc},\text{non}}_{k,\text{thres}}$ as follows,
%
\begin{align}\label{eq_20210311_1}
{q}^{\text{mrc},\text{non}}_{k,\text{thres}}
&\triangleq
Q\left(\left(\frac{\frac{1}{L}\sum\limits^{L}_{j=1}
V_{k,\text{non}}^{\text{mrc}}(j)}
{LT_{c}}\right)^{-\frac{1}{2}}\right.\notag\\
&\left.\quad\times\left(\frac{\ln M_{k}}{LT_{c}}-\frac{\ln\mu}{LT_{c}}
-\frac{1}{L}
\sum\limits^{L}_{j=1}
I_{k,\text{non}}^{\text{mrc},\text{im}}(j)\right)\right).
\end{align}
%
When the channel perturbation ${U}^{\text{mrc}}_{k,\text{non}}$ (given later in \eqref{eq_20210311_7}) satisfies the condition ${U}^{\text{mrc}}_{k,\text{non}}\leq {q}^{\text{mrc},\text{non}}_{k,\text{thres}}$, the achievable coding rate (in nats per channel use) can be expressed as \eqref{eq_20210311_2} shown at the top of next page.
%
\begin{figure*}[!ht]
\begin{align}
\label{eq_20210311_2}
\!\!\!\!\!
\frac{\ln M_{k}}{LT_{c}}
\!\geq\!
R^{\text{mrc}}_{k,\text{non}}
\!\triangleq\! \frac{1}{L}
\sum\limits^{L}_{j=1}I_{k,\text{non}}^{\text{mrc}}(j)
\!+\!\frac{\ln\mu}{LT_{c}}
\!-\!\sqrt{\frac{\frac{1}{L}\sum\limits^{L}_{j=1}
{V}_{k,\text{non}}^{\text{mrc}}(j)}{LT_{c}}}
Q^{-1}\left(\epsilon^{\text{mrc}}_{k,\text{non}}\right)
\!+\!\sqrt{\frac{\frac{1}{L}\sum\limits^{L}_{j=1}
{V}_{k,\text{non}}^{\text{mrc}}(j)}{LT_{c}}}
\mathcal{O}\left({U}^{\text{mrc}}_{k,\text{non}}\right).
\end{align}
\center\hrulefill
\vspace{-2mm}
\end{figure*}
In \eqref{eq_20210311_1} and \eqref{eq_20210311_2}, the parameter $\mu$ is also uniformly distributed on the interval $(0,1)$.
The channel capacity $I_{k,\text{non}}^{\text{mrc}}(j)$ at the $j$-th block, is given by
%
\begin{align}\label{eq_20210311_3}
I_{k,\text{non}}^{\text{mrc}}(j)
=
\frac{T_{c}-\tau_{c}}{T_{c}}\ln\left(1+
{\rm SINR}_{k,\text{non}}^{{\text{mrc}}}(j)\right).
\end{align}
%
The channel dispersion   ${V}_{k,\text{non}}^{\text{mrc}}(j)$, which comes from the variance of the information density, is
%
\begin{align}\label{eq_20210311_5}
{V}_{k,\text{non}}^{\text{mrc}}(j)
&=\frac{(T_{c}-\tau_{c})^{2}}{T_{c}}\left\{\left(\frac{1}{\beta^{{\text{mrc}}}_{y,\text{non}}(j)}
+\frac{1}{\beta^{{\text{mrc}}}_{n,\text{non}}(j)}
\right)\right.\notag\\
&\left.\quad\quad-\frac{2\rho_{\text{non}}^{\text{mrc}}(j)}{\sqrt{\beta^{{\text{mrc}}}_{y,\text{non}}(j)
\beta^{{\text{mrc}}}_{n,\text{non}}(j)}}\right\}.
\end{align}
%
The channel perturbation term ${U}^{\text{mrc}}_{k,\text{non}}$ is given by
%
\begin{align}\label{eq_20210311_7}
{U}^{\text{mrc}}_{k,\text{non}}
=c_{1}\frac{1}{\sqrt{L}}
\left(\frac{T_{c}}{L}\sum\limits^{L}_{j=1}
{V}_{k,\text{non}}^{\text{mrc}}(j)
\right)^{-\frac{3}{2}}
\left(\frac{1}{L}
\sum\limits^{L}_{j=1}
\tilde{U}_{k,\text{non}}^{\text{mrc}}(j)\right),
\end{align}
%
where the term $\tilde{U}_{k,\text{non}}^{\text{mrc}}(j)$ in \eqref{eq_20210311_7}, which represents the absolute third-order moment of the corresponding information density at the $j$-th block, can be approximated by \eqref{eq_20210311_7_1} shown at the top of this page,
%
\begin{figure*}[!ht]
\begin{align}\label{eq_20210311_7_1}
\tilde{U}_{k,\text{non}}^{\text{mrc}}(j)
&\approx
\frac{\left[l^{\text{mrc}}_{22,\text{non}}(j)
\right]^{3+\beta^{{\text{mrc}}}_{\min,\text{non}}(j)}
\left[l^{\text{mrc}}_{11,\text{non}}(j)
-l^{\text{mrc}}_{21,\text{non}}(j)
\right]^{3+l^{\text{mrc}}_{12,\text{non}}(j)}}
{\left[l^{\text{mrc}}_{\text{non}}(j)\right]
^{3+l^{\text{mrc}}_{12,\text{non}}(j)
+\beta^{{\text{mrc}}}_{\min,\text{non}}(j)}}
\notag\\
&\quad\times \left[\beta^{{\text{mrc}}}_{\min,\text{non}}(j)
\right]^{\frac{\beta^{{\text{mrc}}}_{\min,\text{non}}(j)}{2}}
\left[l^{\text{mrc}}_{12,\text{non}}(j)
\right]^{\frac{l^{\text{mrc}}_{12,\text{non}}(j)}{2}}
\frac{\Gamma\left(3+l^{\text{mrc}}_{12,\text{non}}(j)
+\beta^{{\text{mrc}}}_{\min,\text{non}}(j)\right)}
{\Gamma\left(l^{\text{mrc}}_{12,\text{non}}(j)\right)
\Gamma\left(\beta^{{\text{mrc}}}_{\min,\text{non}}(j)\right)}
\Omega_{k,\text{non}}^{\text{mrc}}(j),
\end{align}
\center\hrulefill
\vspace{-2mm}
\end{figure*}
where the term $\Omega_{k,\text{non}}^{\text{mrc}}(j)$ is given by \eqref{eq-2021-11-19-1} shown at the top of this page.
%
\begin{figure*}[!ht]
\begin{align}\label{eq-2021-11-19-1}
\Omega_{k,\text{non}}^{\text{mrc}}(j)
&=\sum\limits^{3}_{f=0}\frac{(-1)^{f}3!}{f!(3-f)!}
\left\{
\frac{{}_{2}F_{1}\left(1,3+l^{\text{mrc}}_{12,\text{non}}(j)
+\beta^{{\text{mrc}}}_{\min,\text{non}}(j),
f+\beta^{{\text{mrc}}}_{\min,\text{non}}(j)+1;
\frac{l^{\text{mrc}}_{22,\text{non}}(j)
\sqrt{\beta^{{\text{mrc}}}_{\min,\text{non}}(j)}}
{l^{\text{mrc}}_{\text{non}}(j)}
\right)}{f+\beta^{{\text{mrc}}}_{\min,\text{non}}(j)}\right.\notag\\
&\left.\quad\quad
+\frac{{}_{2}F_{1}\left(1,3+l^{\text{mrc}}_{12,\text{non}}(j)
+\beta^{{\text{mrc}}}_{\min,\text{non}}(j),
f+l^{\text{mrc}}_{12,\text{non}}(j)+1;
\frac{\left[l^{\text{mrc}}_{11,\text{non}}(j)
-l^{\text{mrc}}_{21,\text{non}}(j)\right]
\sqrt{l^{\text{mrc}}_{12,\text{non}}(j)}}
{l^{\text{mrc}}_{\text{non}}(j)}\right)}
{f+l^{\text{mrc}}_{12,\text{non}}(j)}\right\}.
\end{align}
\center\hrulefill
\vspace{-3mm}
\end{figure*}
In \eqref{eq_20210311_5}-\eqref{eq-2021-11-19-1}, the parameters
$\rho_{\text{non}}^{\text{mrc}}(j)$, $\beta^{{\text{mrc}}}_{y,\text{non}}(j)$, and $\beta^{{\text{mrc}}}_{n,\text{non}}(j)$,
have been given in  \eqref{eq_20210930_1}, \eqref{eq_im_mrc_20210309_20},  and \eqref{eq_im_mrc_20210312_2}, respectively. Note that the hypergeometric function ${}_{2}F_{1}\left(\cdot\right)$ in \eqref{eq-2021-11-19-1} is convergent.
\end{prop}
\begin{IEEEproof}
See Appendix \ref{appendix_prop_imper_mrc}.
\end{IEEEproof}
\begin{remark}
\label{remark_mrc_non_coherent}
Note that the channel dispersion ${V}_{k,\text{non}}^{\text{mrc}}(j)$ of \eqref{eq_20210311_5} is a closed-form result while $\tilde{U}_{k,\text{non}}^{\text{mrc}}(j)$ in \eqref{eq_20210311_7_1} is obtained based on the matched Gamma approximations in \eqref{eq_v_yn_gamma} \& \eqref{eq_v_yn_gamma-n}
and the absolute high-order moment approximation in Lemma \ref{lemma_20210531_1} of Appendix \ref{preliminary}.
From the simulations later, the approximation result of \eqref{eq_20210311_7_1} is quite accurate and the gap from its real value can be almostly ignored. The channel dispersion ${V}_{k,\text{non}}^{\text{mrc}}(j)$ and channel perturbation ${U}^{\text{mrc}}_{k,\text{non}}$ of the achievable coding rate in the non-coherent case are acquired by fixing $s=\frac{1}{\sigma^{2}_{\check{{\bf n}}^{\text{mrc}}_{\text{non}}}(j)}$, since this is a capacity achieving strategy.
Now, comparing Proposition \ref{prop_imper_mrc} with  Proposition \ref{prop_20200103_1}, we have:
\begin{enumerate}[i)]
\item Firstly, the expressions in \eqref{eq_20210311_1}-\eqref{eq_20210311_2} in Proposition \ref{prop_imper_mrc} are similar as the results given in Proposition \ref{prop_20200103_1} of Section \ref{mrc_scheme_trans}. Thus, the same conclusions as in Remark \ref{remark_mrc_per} can also be obtained for the non-coherent scenario;
\item Secondly,
the channel capacity $I_{k,\text{non}}^{\text{mrc}}(j)$ in \eqref{eq_20210311_3} for a given $j$-th block and the corresponding average result under block fading channels (the first term in \eqref{eq_20210311_2}) are determined only by the ${\rm SINR}_{k,\text{non}}^{{\text{mrc}}}(j)$, for $j=1,
\ldots,L$. This is the same as the MRC scheme in the coherent case. However, the channel dispersion ${V}_{k,\text{non}}^{\text{mrc}}(j)$ in \eqref{eq_20210311_5} and the perturbation
${U}^{\text{mrc}}_{k,\text{non}}$ in \eqref{eq_20210311_7} are determined not only by the ${\rm SINR}_{k,\text{non}}^{{\text{mrc}}}(j)$, but also by the parameters $\frac{P_{k}(\gamma^{2}_{k}-\phi^{2}_{k})}
{\sigma^{2}_{\check{{\bf n}}_{\text{non}}^{\text{mrc}}}(j)}$ and $\frac{\Delta^{\text{mrc}}_{k,\text{non}}(j)}{\sigma^{4}_{\check{{\bf n}}_{\text{non}}^{\text{mrc}}}(j)}$, which are brought in by the channel estimation errors from all users and represent the exact difference from the corresponding coherent case. Thus, setting all  channel estimation errors $\gamma^{2}_{m}-\phi^{2}_{m}=0$, for all $m=1,\ldots, K_{u}$, from the matched Gamma approximation in \eqref{eq_v_yn_gamma} and the related parameters in \eqref{eq_20210930_1}-\eqref{eq_20210311_6},
the non-coherent case simplifies to the corresponding coherent case.
\end{enumerate}
\end{remark}

The channel dispersion and perturbation terms in Proposition \ref{prop_imper_mrc} are quite general and valid for the system with any arbitrary and finite number of BS antennas. 
However, these expressions are quite complicated.
In the following, an asymptotic analysis will be
performed. Again, by using the law of large numbers,
the related parameters in Proposition \ref{prop_imper_mrc}, which includes ${\rm SINR}_{k,\text{non}}^{{\text{mrc}}}(j)$ of \eqref{sinr_mrc_im}, the noise variance $\sigma^{2}_{\check{{\bf n}}^{\text{mrc}}_{\text{non}}}(j)$ of \eqref{eq_im_20210307_4} and the parameter $\Delta^{\text{mrc}}_{k,\text{non}}(j)$ of \eqref{eq_20210311_6},
will approach to
%
\begin{align}\label{eq_20210312_5}
{\rm SINR}_{k,\text{non}}^{{\text{mrc}}}(j)
\xrightarrow[]{N_{b} \rightarrow \infty}\infty,
\end{align}
%
\begin{align}\label{eq_20210312_18}
\sigma^{2}_{\check{{\bf n}}^{\text{mrc}}_{\text{non}}}(j)
&\xrightarrow[]{N_{b} \rightarrow \infty}
\bar{\sigma}^{2}_{\check{{\bf n}}^{\text{mrc}}_{\text{non}}}\notag\\
&=P_{k}\left(\gamma^{2}_{k}-\phi_{k}^2\right)
+\sum\limits^{K_{u}}_{\substack{m=1\\ m\neq k}}P_{m}\gamma^{2}_{m}
+\sigma^{2}_{n},
\end{align}
%
\begin{align}\label{eq_20210312_19}
\Delta^{\text{mrc}}_{k,\text{non}}(j)
&\xrightarrow[]{N_{b} \rightarrow \infty}
\bar{\Delta}^{\text{mrc}}_{k,\text{non}}\notag\\
&=P^{2}_{k}(\gamma^{2}_{k}-\phi^{2}_{k})^{2}
+\sum\limits^{K_{u}}_{\substack{m=1\\ m\neq k}}
P^{2}_{m}(\gamma^{4}_{m}-\phi^{4}_{m}).
\end{align}
%
Substituting these asymptotic results of \eqref{eq_20210312_5}-\eqref{eq_20210312_19}
into \eqref{eq_20210930_1}-\eqref{eq_im_mrc_20210312_2},
we further have:
%
\begin{align}\label{eq_20210930_3}
\beta^{{\text{mrc}}}_{y,\text{non}}(j)
\xrightarrow[]{N_{b} \rightarrow \infty}
\bar{\beta}^{{\text{mrc}}}_{y,\text{non}}
=T_{c}-\tau_{c},
\end{align}
%
\begin{align}\label{eq_20210312_10}
&\beta^{{\text{mrc}}}_{n,\text{non}}(j)
\xrightarrow[]{N_{b} \rightarrow \infty}
\bar{\beta}^{{\text{mrc}}}_{n,\text{non}}
=\frac{T_{c}-\tau_{c}}{1+\left[(T_{c}-\tau_{c})+1\right]
\frac{\bar{\Delta}^{\text{mrc}}_{k,\text{non}}}
{\bar{\sigma}^{4}_{\check{{\bf n}}^{\text{mrc}}_{\text{non}}}}},
\end{align}
%
\begin{align}\label{eq_20210930_4}
\rho_{\text{non}}^{\text{mrc}}(j)
\xrightarrow[]{N_{b} \rightarrow \infty}
\bar{\rho}_{\text{non}}^{\text{mrc}}
=\frac{P_{k}(\gamma^{2}_{k}-\phi^{2}_{k})}
{\sqrt{\bar{\sigma}^{4}_{\check{{\bf n}}^{\text{mrc}}_{\text{im}}}
+\left[(T_{c}-\tau_{c})+1\right]
\bar{\Delta}^{\text{mrc}}_{k,\text{non}}}}.
\end{align}
%

Substituting \eqref{eq_20210312_5}-\eqref{eq_20210930_4} into
Proposition \ref{prop_imper_mrc}, the following corollary can be obtained.
\begin{corollary}
\label{coro_imper_mrc}
Without considering any power scaling laws, when $N_{b}\rightarrow \infty$,
the channel capacity
$I_{k,\text{non}}^{\text{mrc}}(j)$ in \eqref{eq_20210311_3}, the channel dispersion ${V}_{k,\text{non}}^{\text{mrc}}(j)$ in \eqref{eq_20210311_5} and the channel  perturbation  term ${U}^{\text{mrc}}_{k,\text{non}}$ in \eqref{eq_20210311_7}  will approach to
%
\begin{align}\label{eq_20210312_12}
I_{k,\text{non}}^{\text{mrc}}(j)
\xrightarrow[]{N_{b} \rightarrow \infty}
\infty,
\end{align}
%
\begin{align}\label{eq_20210312_13}
{V}_{k,\text{non}}^{\text{mrc}}(j)
&\xrightarrow[]{N_{b} \rightarrow \infty}
\bar{{V}}_{k,\text{non}}^{\text{mrc}}\notag\\
&=\frac{(T_{c}-\tau_{c})^{2}+(T_{c}-\tau_{c})}{T_{c}}
\frac{\bar{\Delta}^{\text{mrc}}_{k,\text{non}}}
{\bar{\sigma}^{4}_{\check{{\bf n}}^{\text{mrc}}_{\text{non}}}}\notag\\
&\quad+\frac{(T_{c}-\tau_{c})}{T_{c}}
\left(2-2\frac{P_{k}(\gamma^{2}_{k}-\phi_{k}^2)}
{\bar{\sigma}^{2}_{\check{{\bf n}}^{\text{mrc}}_{\text{non}}}}\right),
\end{align}
%
\begin{align}\label{eq_20210312_14}
{U}^{\text{mrc}}_{k,\text{non}}
\xrightarrow[]{N_{b} \rightarrow \infty}
\bar{{U}}^{\text{mrc}}_{k,\text{non}}
=\frac{c_{1}}{\sqrt{L}}
\left(T_{c}\bar{{V}}_{k,\text{non}}^{\text{mrc}}
\right)^{-\frac{3}{2}}
\bar{\tilde{{U}}}^{\text{mrc}}_{k,\text{non}},
\end{align}
%
where $\bar{\tilde{{U}}}^{\text{mrc}}_{k,\text{non}}$ is the asymptotic result of
$\tilde{{U}}^{\text{mrc}}_{k,\text{non}}(j)$ in \eqref{eq_20210311_7}, which can be obtained by directly replacing  $\beta^{{\text{mrc}}}_{y,\text{non}}(j)$, $\beta^{{\text{mrc}}}_{n,\text{non}}(j)$,
$\rho_{\text{non}}^{\text{mrc}}(j)$ in \eqref{eq_20210311_7_1} and \eqref{eq-2021-11-19-1} with $\bar{\beta}^{{\text{mrc}}}_{y,\text{non}}$, $\bar{\beta}^{{\text{mrc}}}_{n,\text{non}}$, $\bar{\rho}_{\text{non}}^{\text{mrc}}$ of
\eqref{eq_20210930_3}-\eqref{eq_20210930_4}.
In \eqref{eq_20210312_13}, the parameters $\bar{\sigma}^{2}_{\check{{\bf n}}^{\text{mrc}}_{\text{non}}}$ and  $\bar{\Delta}^{\text{mrc}}_{k,\text{non}}$ have been given in \eqref{eq_20210312_18} and \eqref{eq_20210312_19}, respectively.
\end{corollary}
\begin{remark}\label{remark_im_mrc}
As the channel dispersion $\bar{{V}}_{k,\text{non}}^{\text{mrc}}$ in
\eqref{eq_20210312_13} and  perturbation term  $\bar{{U}}^{\text{mrc}}_{k,\text{non}}$ in \eqref{eq_20210312_14} are constants and independent of $N_{b}$, the conclusions obtained in
Remark \ref{remark_0810_1} of Section \ref{mrc_scheme_trans} for the MRC scheme  also hold here. Meanwhile,
from Corollary \ref{coro_imper_mrc}, we have that: under our system setting and in the finite blocklength regime, without considering any power scaling laws, massive MIMO fails in eliminating the channel estimation errors in the channel dispersion and perturbation term, which is quite different from the usual infinite blocklength case. This reveals that imperfect CSI is much more harmful in the finite blocklength regime even in the massive MIMO region. The reason behind this is that the channel estimation errors in the asymptotic results $\bar{\rho}_{\text{non}}^{\text{mrc}}$ and $\bar{\beta}^{{\text{mrc}}}_{n,\text{non}}$ still exist in the massive MIMO regime. Recall the physical meanings of these two terms in \eqref{eq_20210930_1} and \eqref{eq_im_mrc_20210312_2},
we can conclude that in the non-coherent case,  the  effective noise cannot be treated as Gaussian distributed and the output signal and  noise vector cannot be assumed to be uncorrelated unless the channel estimation error from the intended user $k$ is equal to zero, i.e., $\gamma^{2}_{k}-\phi^{2}_{k}=0$. This is in sharp contrast to the corresponding results obtained in Corollary \ref{cor_0824_1} and
this key observation will be verified clearly via simulations.
\end{remark}

We will now perform the asymptotic analysis under the power scaling laws at both the pilot transmission phase and signal transmission phase.
With this, we have the following corollary.
\begin{corollary}
\label{coro_asy_non_mrc_power}
When the power scaling law
$P_{k}=\frac{E_{k}}{\sqrt{N_{b}}}$
and $P^{c}_{k}=\frac{E^{c}_{k}}{\sqrt{N_{b}}}$ with $E_{k}$ and $E^{c}_{k}$ fixed, for all $k=1,\ldots,K_{u}$, is considered,  we have,
%
\begin{align}\label{eq-2021-11-09-6}
{\rm SINR}_{k,\text{non}}^{{\text{mrc}}}(j)
\xrightarrow[]{N_{b} \rightarrow \infty}
\overline{\rm SINR}_{k,\text{non}}^{{\text{mrc}},\infty}
=
\frac{\tau_{c}\gamma^{4}_{k}}
{\sigma^{2}_{c}\sigma^{2}_{n}}E_{k}E^{c}_{k},
\end{align}
%
while the parameters $\beta^{{\text{mrc}}}_{y,\text{non}}(j)$, $\beta^{{\text{mrc}}}_{n,\text{non}}(j)$, and $\rho_{\text{non}}^{\text{mrc}}(j)$ in \eqref{eq_im_mrc_20210309_20}, \eqref{eq_im_mrc_20210312_2}, and \eqref{eq_20210930_1}
will respectively approach to
%
\begin{align}
&\beta^{{\text{mrc}}}_{y,\text{non}}(j),
\beta^{{\text{mrc}}}_{n,\text{non}}(j)
\xrightarrow[]{N_{b} \rightarrow \infty}
T_{c}-\tau_{c},\label{eq-2021-11-09-7}\\
%
&\rho_{\text{non}}^{\text{mrc}}(j)
\xrightarrow[]{N_{b} \rightarrow \infty}
\frac{1}{1+\overline{\rm SINR}_{k,\text{non}}^{{\text{mrc}},\infty}}.
\label{eq-2021-11-09-7-rho}
\end{align}
\end{corollary}
\begin{remark}\label{remark-non-power-mrc-1}
Comparing the asymptotic results in  Corollary \ref{coro_asy_non_mrc_power} with those in Corollary \ref{coro_imper_mrc},
it is clear that under the power scaling laws given in Corollary \ref{coro_asy_non_mrc_power}, all the channel estimation errors are theoretically eliminated with a massive number of antennas. From the asymptotic results
 $\beta^{{\text{mrc}}}_{y,\text{non}}(j)$ and
$\beta^{{\text{mrc}}}_{n,\text{non}}(j)$ obtained in \eqref{eq-2021-11-09-7}, both the  effective noise vector $\check{{\bf n}}^{\text{mrc}}_{\text{non}}(j)$ in \eqref{eq_im_20210307_1} and the output signal vector
$\tilde{{\bf y}}^{\text{mrc}}_{k,\text{non}}(j)$ in \eqref{eq_im_20210307_2} would approach to be Gaussian distributed. However, from the asymptotic result of $\rho_{\text{non}}^{\text{mrc}}(j)$ in \eqref{eq-2021-11-09-7-rho}, the  effective noise vector $\check{{\bf n}}^{\text{mrc}}_{\text{non}}(j)$ and the output signal vector
$\tilde{{\bf y}}^{\text{mrc}}_{k,\text{non}}(j)$ will never be independent under the scenario considered in Corollary \ref{coro_asy_non_mrc_power}. These insights are in sharp contrast to the results obtained in Corollary \ref{coro_imper_mrc} but similar with the results obtained in Corollary \ref{corollary_co_mrc_power}.

The asymptotic results in Corollary \ref{coro_asy_non_mrc_power} are very similar with  the general case in Proposition \ref{prop_20200103_1}  for the MRC scheme in the coherent CSI in Section \ref{mrc_scheme_trans}.
Thus, we have the following corollary.
\end{remark}
\begin{corollary}
\label{coro_asy_non_mrc_power_ca_dis}
Under the power scaling laws $P_{k}=\frac{E_{k}}{\sqrt{N_{b}}}$
and $P^{c}_{k}=\frac{E^{c}_{k}}{\sqrt{N_{b}}}$, for all $k=1,\ldots,K_{u}$, and massive MIMO at the BS, the channel capacity
$I_{k,\text{non}}^{\text{mrc}}(j)$ in \eqref{eq_20210311_3}, the channel dispersion ${V}_{k,\text{non}}^{\text{mrc}}(j)$ in \eqref{eq_20210311_5} and the  perturbation  term ${U}^{\text{mrc}}_{k,\text{non}}$ in \eqref{eq_20210311_7} of Proposition \ref{prop_imper_mrc} will approach to
%
\begin{align}
\label{eq-2021-11-09-8}
I_{k,\text{non}}^{\text{mrc}}(j)
&\xrightarrow[]{N_{b} \rightarrow \infty}
\bar{\bar{I}}_{k,\text{non}}^{\text{mrc},\infty}\notag\\
&=\frac{T_{c}-\tau_{c}}{T_{c}}
\ln\left(1+\frac{\tau_{c}\gamma^{4}_{k}}
{\sigma^{2}_{c}\sigma^{2}_{n}}E_{k}E^{c}_{k}\right),
\end{align}
%
\begin{align}
\label{eq-2021-11-09-9}
{V}_{k,\text{non}}^{\text{mrc}}(j)
&\xrightarrow[]{N_{b} \rightarrow \infty}
\bar{\bar{V}}_{k,\text{non}}^{\text{mrc},\infty}\notag\\
&=\frac{T_{c}-\tau_{c}}{T_{c}}
\left(2-\frac{2}{1+\frac{\tau_{c}\gamma^{4}_{k}}
{\sigma^{2}_{c}\sigma^{2}_{n}}E_{k}E^{c}_{k}}\right),
\end{align}
%
\begin{align}
\label{eq-2021-11-09-10}
{U}^{\text{mrc}}_{k,\text{non}}
&\xrightarrow[]{N_{b} \rightarrow \infty}
\bar{{U}}^{\text{mrc},\infty}_{k,\text{non}}\notag\\
&=\frac{4c_{1}}{\sqrt{2\pi}}
\frac{1}{\sqrt{L}}
\frac{1}{\sqrt{T_{c}-\tau_{c}}}\notag\\
&\quad\quad\times\frac{1}{\Gamma\left(T_{c}-\tau_{c}+1\right)}
\Gamma\left(\frac{2(T_{c}-\tau_{c})+3}{2}\right).
\end{align}
\end{corollary}
\begin{remark}
\label{remark-2021-11-10-1}
The results obtained in Corollary \ref{coro_asy_non_mrc_power_ca_dis} are quite similar with those obtained in Corollary \ref{corollary_co_mrc_power} for the MRC scheme with the  power scaling law $P_{k}=\frac{E_{k}}{N_{b}}$ under the coherent case, except the $\tau_{c}$ and asymptotic {\rm SINR} difference. Hence, the corresponding conclusions in Remark \ref{remark_co_mrc_power_scaling} are also valid here.
Furthermore,
by using the result in \cite{hoydis2015second}, we have
%
\begin{align}
\label{eq-2021-11-09-1}
&\quad\frac{4c_{1}}{\sqrt{2\pi}}
\frac{1}{\sqrt{L}}
\frac{T_{c}-\tau_{c}+1}{\sqrt{\left(T_{c}-\tau_{c}\right)
\left(T_{c}-\tau_{c}+3/2\right)}}\notag\\
&\leq
\bar{{U}}^{\text{mrc},\infty}_{k,\text{non}}
\leq \frac{4c_{1}}{\sqrt{2\pi}}
\frac{1}{\sqrt{L}}\left(1+\frac{1}{T_{c}-\tau_{c}}
\right)^{1/2}.
\end{align}
%
Both the left-hand and right-hand side of \eqref{eq-2021-11-09-1} are decreasing functions with respect to $T_{c}-\tau_{c}$. This indicates that in the asymptotic scene, using longer pilot sequences  is not beneficial for suppressing the perturbation term of the achievable coding rate in the non-coherent case. Meanwhile, the channel dispersion $\bar{\bar{V}}_{k,\text{non}}^{\text{mrc},\infty}$ in \eqref{eq-2021-11-09-9} is a concave function with respect to the pilot length $\tau_{c}$.
\end{remark}
%
%
\subsection{ZF Scheme}\label{zf_im}
With the estimated channels in \eqref{eq_im_20210306_1}, the ZF combining vector of \eqref{eq_20200106_1} will become
%
\begin{align}\label{eq_zf_20210312_1}
{\bf a}^{\text{zf}}_{k,\text{non}}(j)
=\frac{\left(\hat{{\bf h}}^{\dagger}_{k}(j)\hat{{\bf V}}_{k}(j)\right)^{\dagger}}{\|\hat{{\bf h}}^{\dagger}_{k}(j)\hat{{\bf V}}_{k}(j)\|_{2}},
\end{align}
%
where $\hat{{\bf V}}_{k}(j)$ has similar definition as \eqref{eq_zf_20210312_2} by replacing $\{{\bf h}_{k}(j)\}^{K_{u}}_{k=1}$ with the  estimated channels $\{\hat{{\bf h}}_{k}(j)\}^{K_{u}}_{k=1}$ of \eqref{eq_im_20210306_1}.
Now, substituting \eqref{eq_zf_20210312_1} and the estimated channel of \eqref{eq_im_20210306_1} into  \eqref{eq_20191015_1} yields \eqref{eq_zf_20210312_3} shown at the top of this page,
%
\begin{figure*}[!ht]
\begin{align}\label{eq_zf_20210312_3}
\tilde{{\bf y}}^{\text{zf}}_{k,\text{non}}(j)
=\|\hat{{\bf h}}^{\dagger}_{k}(j)\hat{{\bf V}}_{k}(j)\|_{2}{\bf x}^{T}_{k}(j)
+\underbrace{\frac{\hat{{\bf h}}^{\dagger}_{k}(j)\hat{{\bf V}}_{k}(j)}{\|\hat{{\bf h}}^{\dagger}_{k}(j)\hat{{\bf V}}_{k}(j)\|_{2}}\tilde{{\bf H}}(j)({\bf X}(j))^{T}
\!+\!\frac{\hat{{\bf h}}^{\dagger}_{k}(j)\hat{{\bf V}}_{k}(j)}{\|\hat{{\bf h}}^{\dagger}_{k}(j)\hat{{\bf V}}_{k}(j)\|_{2}}{\bf N}(j)}_{\triangleq \check{{\bf n}}^{\text{zf}}_{\text{non}}(j)},
\end{align}
\vspace{-5mm}
\center\hrulefill
\vspace{-3mm}
\end{figure*}
where $\tilde{{\bf H}}(j)=[\tilde{\bf h}_{1}(j),\ldots,\tilde{\bf h}_{K_{u}}(j)]\in\mathcal{C}^{N_{b}\times K_{u}}$. 
The term $\check{{\bf n}}^{\text{zf}}_{\text{non}}(j)\in \mathcal{C}^{1\times (T_{c}-\tau_{c})}$ denotes the  effective noise vector, which includes independent non-Gaussian random elements and each element has zero mean and the variance
%
\begin{align}\label{eq_im_zf_20210312_6}
\sigma^{2}_{\check{{\bf n}}^{\text{zf}}_{\text{non}}}
=\sum\limits^{K_{u}}_{k=1}P_{k}(\gamma^{2}_{k}-\phi_{k}^2)
+\sigma^{2}_{n}.
\end{align}
%
Hence, \eqref{eq_zf_20210312_3} can also be treated as a point-to-point SISO channel with the i.i.d. Gaussian input signal vector ${\bf x}^{T}_{k}(j)\in \mathcal{C}^{1\times (T_{c}-\tau_{c})}$, fading channel $\|\hat{{\bf h}}^{\dagger}_{k}(j)\hat{{\bf V}}_{k}(j)\|_{2}$ and the non-Gaussian noise vector $\check{{\bf n}}^{\text{zf}}_{\text{non}}(j)$. The corresponding signal-to-interference-plus-noise-ratio ${\rm SINR}_{k,\text{non}}^{{\text{zf}}}(j)$ at the $j$-th block for the non-coherent and ZF case is
%
\begin{align}\label{sinr_zf_im}
{\rm SINR}_{k,\text{non}}^{{\text{zf}}}(j)
=\frac{P_{k}\|\hat{{\bf h}}^{\dagger}_{k}(j)\hat{{\bf V}}_{k}(j)\|^{2}_{2}}{\sigma^{2}_{\check{{\bf n}}^{\text{zf}}_{\text{non}}}}.
\end{align}

Note that this case is similar with the MRC scheme for the non-coherent setting in Section \ref{section_im_mrc}. Thus, we define the coefficient $\Delta^{\text{zf}}_{k,\text{non}}$ for the ZF scheme, which is as
\begin{align}\label{eq_im_zf_20210312_16}
\Delta^{\text{zf}}_{k,\text{non}}
=\sum\limits^{K_{u}}_{k=1}
P^{2}_{k}(\gamma^{2}_{k}-\phi^{2}_{k})^{2}.
\end{align}
%
By replacing the parameters ${\rm SINR}_{k,\text{non}}^{{\text{mrc}}}(j)$, $\sigma^{2}_{\check{{\bf n}}^{\text{mrc}}_{\text{non}}}$, and $\Delta^{\text{mrc}}_{k,\text{non}}$ that are included in the terms $\beta^{{\text{mrc}}}_{y,\text{non}}(j)$, $\beta^{{\text{mrc}}}_{n,\text{non}}(j)$, and $\rho_{\text{non}}^{\text{mrc}}(j)$ in Proposition \ref{prop_imper_mrc} with that ${\rm SINR}_{k,\text{non}}^{{\text{zf}}}(j)$ of \eqref{sinr_zf_im}, $\sigma^{2}_{\check{{\bf n}}^{\text{zf}}_{\text{non}}}$ of \eqref{eq_im_zf_20210312_6}, and $\Delta^{\text{zf}}_{k,\text{non}}$ of \eqref{eq_im_zf_20210312_16}, respectively, similar conclusions about the achievable coding rate as given in Proposition \ref{prop_imper_mrc} can also be obtained for the ZF scheme.  Meanwhile, by replacing the asymptotic parameters $\bar{\sigma}^{2}_{\check{{\bf n}}^{\text{mrc}}_{\text{non}}}$ and $\bar{\Delta}^{\text{mrc}}_{k,\text{non}}$ included in all the terms of Corollary \ref{coro_imper_mrc} with that $\sigma^{2}_{\check{{\bf n}}^{\text{zf}}_{\text{non}}}$ of \eqref{eq_im_zf_20210312_6} and $\Delta^{\text{zf}}_{k,\text{non}}$ of \eqref{eq_im_zf_20210312_16}, respectively, similar asymptotic results as given in Corollary \ref{coro_imper_mrc} can also be obtained for the ZF scheme.
Furthermore, under the power scaling laws $P_{k}=\frac{E_{k}}{\sqrt{N_{b}}}$
and $P^{c}_{k}=\frac{E^{c}_{k}}{\sqrt{N_{b}}}$, for all $k=1,\ldots,K_{u}$, the same asymptotic result as in Corollary \ref{coro_asy_non_mrc_power_ca_dis} can be obtained here.
\vspace{-3mm}
\section{Packet Loss Probability in the Finite Blocklength Regime}\label{packet_loss_probability}
In the following, we will present the relationship between the packet loss probability \cite{park2020performance} and the average error probability in the finite blocklength regime.
Firstly, in the finite blocklength regime, we consider one packet is lost if the receiver cannot correctly decode all symbols transmitted by the intended user. In this paper, we will mainly consider the following two scenarios  that lead to a packet loss:
\begin{enumerate}[i)]
\item The first one occurs whenever the achievable coding rate is below a target threshold; consequently, the transmitter cannot send information to the receiver successfully due to the poor channel conditions;
\item The second one occurs whenever the channels are good enough for successful transmission, however,  the receiver cannot decode the desired signals completely, i.e., the  decoding error probability is positive. This is quite different from the infinite blocklength regime, in which the decoding error probability can be arbitrarily small.
\end{enumerate}
Thus, combining these two cases, we have the following corollary:
\begin{corollary}\label{theorem-2021-11-10-1}
In the finite blocklength regime, for a fixed average error probability, the packet loss probability for the intended $k$-th user in our paper can be expressed as
%
\begin{align}\label{outage}
{\rm Pr}_{k}^{\text{loss}}
&=\operatorname{Pr}\left\{
R^{\text{app}}_{k}
<\tilde{R}^{\text{thres}}_{k}\right\}
+\epsilon_{k}\operatorname{Pr}\left\{
R^{\text{app}}_{k}
\geq \tilde{R}^{\text{thres}}_{k}\right\}\notag\\
&=\epsilon_{k}+\left(1-\epsilon_{k}\right)
\operatorname{Pr}\left\{
\epsilon_{k}<q^{\text{thres}}_{k,\text{loss}}\right\},
\end{align}
%
where $q^{\text{thres}}_{k,\text{loss}}$ is given by
%
\begin{align}\label{eq_20211105_1}
q^{\text{thres}}_{k,\text{loss}}
&=Q\left(\left(\frac{\frac{1}{L}\sum\limits^{L}_{j=1}
V_{k}(j)}
{LT_{c}}\right)^{-\frac{1}{2}}\right.\notag\\
&\left.\quad\quad\times
\left(\frac{1}{L}
\sum\limits^{L}_{j=1}
I_{k}(j)
+\frac{\ln\mu}{LT_{c}}-\tilde{R}^{\text{thres}}_{k}
\right)\right).
\end{align}
%
In \eqref{outage} and \eqref{eq_20211105_1}, the term $\tilde{R}^{\text{thres}}_{k}$ denotes a predetermined  target rate for the $k$-th user. The parameter $\epsilon_{k}\in \left\{\epsilon^{\text{mrc}}_{k,\text{co}}, \epsilon^{\text{zf}}_{k,\text{co}}, \epsilon^{\text{mrc}}_{k,\text{non}}, \epsilon^{\text{zf}}_{k,\text{non}}\right\}$ represents all the average decoding error probabilities. The term $R^{\text{app}}_{k}
\in \left\{R^{\text{mrc},\text{co}}_{k,\text{app}},
R^{\text{zf},\text{co}}_{k,\text{app}}, R^{\text{mrc},\text{non}}_{k,\text{app}}, R^{\text{zf},\text{non}}_{k,\text{app}}\right\}$ represents all the achievable coding rates obtained in this paper after removing the perturbation term, and the parameters ${V}_{k}\in \left\{{V}_{k,\text{co}}^{\text{mrc}}, {V}_{k,\text{co}}^{\text{zf}}, {V}_{k,\text{non}}^{\text{mrc}}, {V}_{k,\text{non}}^{\text{zf}}\right\}$ and $I_{k}\in\left\{I_{k,\text{co}}^{\text{mrc}}, I_{k,\text{co}}^{\text{zf}}, I_{k,\text{non}}^{\text{mrc}}, I_{k,\text{non}}^{\text{zf}}\right\}$ represent all the channel dispersions and channel capacities, respectively.
\end{corollary}
\begin{remark}\label{remark-2021-11-10-4}
Note that the packet loss probability in Corollary \ref{theorem-2021-11-10-1} is based on the premise that we treat the average decoding error probability caused by the finite blocklength as one of the packet loss events. In this context, from \eqref{outage}, it is easy to obtain that
${\rm Pr}_{k}^{\text{loss}}\geq \epsilon_{k}$,
 i.e., the packet loss probability is lower bounded by the average error probability.
From the asymptotic analysis in this paper,
without considering any power scaling laws, all the channel capacities $I_{k}$ approach to infinity. In this situation, the term $\operatorname{Pr}\left\{
\epsilon_{k}<q^{\text{thres}}_{k,\text{loss}}\right\}$ in \eqref{outage} approaches to zero and
in this context, the packet loss probability  can be approximated by the corresponding average error probability $\epsilon_{k}$  in the massive MIMO regime. This implies that in massive MIMO system without considering any power scaling laws, the packet loss probability tends to coincide with the average error probability. However, under the power scaling laws, the result will be very different due to that the term $q^{\text{thres}}_{k,\text{loss}}$ in \eqref{eq_20211105_1} approaches to a non-zero constant.

On the other hand, Corollary \ref{theorem-2021-11-10-1} can also be explained in the following way: define the effective decoding error probability as
%
${\small\epsilon^{\text{eff}}_{k}
\triangleq
\begin{cases}
1, & R^{\text{app}}_{k}<\tilde{R}^{\text{thres}}_{k} \\
\epsilon_{k}, & R^{\text{app}}_{k}
\geq \tilde{R}^{\text{thres}}_{k}
\end{cases}}$;
then, the packet loss probability in the finite blocklength regime is defined as ${\rm Pr}_{k}^{\text{loss}}
=\mathbb{E}\left\{\epsilon^{\text{eff}}_{k}\right\}$.
\end{remark}
%
%
\section{Numerical Results}\label{numerical}
%
\subsection{Parameters Setting}
We consider a single cell with radius $r=300$m and all users are uniformly distributed in this cell.
The simulations are carried out under some basic system parameters setting unless otherwise stated. The number of the BS antennas is set as $N_{b}=20$, whilst
the number of users is $K_{u}=5$ and the pilot length is $\tau_{c}=5$.
The transmit powers in both the channel estimation phase and signal transmission phase are $P_{k}=P^{c}_{k}=P=10$dBm, for $k=1,\ldots,K_{u}$, and the noise variance are set as $\sigma_{n}^{2}=\sigma_{c}^{2}=-100$dBm.  The coherence time is $T_{c}=20$ while the number of blocks is $L=10$. We assume
the same information nats for all users that allowed to be transmitted are as $\ln M_{1}=\ldots=\ln M_{K_{u}}=b=100$nats.
The average error probabilities are $\epsilon_{1}=\ldots=\epsilon_{K_{u}}=\epsilon=10^{-5}$.
The large-scale fading coefficients $\gamma^{2}_{k}$, $k=1,\ldots,K_{u}$, can be modeled in dB as \cite{bjornson2017massive}
%
\begin{align}\label{eq_20210503_1}
(\gamma^{2}_{k})_{\text{dB}}
=-35.3-37.6\log_{10}\left(d_{k}\right)
+z_{k},
\end{align}
%
where $d_{k}$ in meters is the distance from the $k$-th user to the BS. The non-deterministic variable $z_{k}$ is called the shadow fading and distributed as $z_{k}\sim \mathcal{N}\left(0,\sigma^{2}_{z}\right)$, in which $\sigma_{z}=8$dB is the standard deviation of the log-normal random variable. Note that all the simulation results are presented without considering the power scaling laws due to limited space.
\vspace{-3mm}
\subsection{High-order Moments and the Matched Gamma Approximation}
\begin{figure*}[t!]
  \advance\leftskip-1.7cm
  \includegraphics
  [width=21cm,
  height=5.0cm]
  {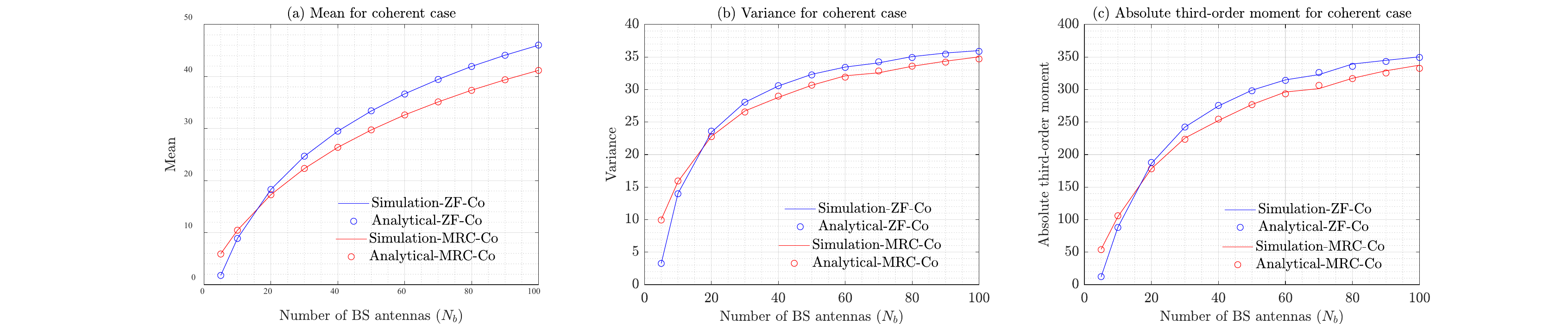}
  \caption{The high-order moments of the information density for a given  block for both MRC and ZF schemes versus $N_{b}$ with $T_{c}=20$ (coherent case).}\label{fig_1}
\end{figure*}

\begin{figure*}[t!]
  \advance\leftskip-1.7cm
  \includegraphics
  [width=21.0cm,height=4.5cm]
  {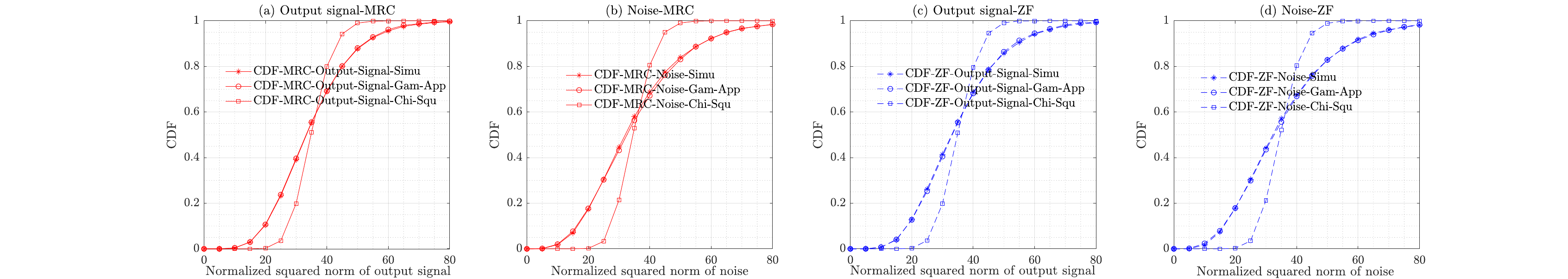}
  \caption{The simulation and matched Gamma approximation for the normalized squared norm of both the output signal and noise vector under MRC and ZF schemes with $N_{b}=20$, $T_{c}=40$, $P_{k}=40$dBm, and $L=10$ (non-coherent case).}\label{fig_im_CDF}
\end{figure*}

\begin{figure*}[t!]
  \advance\leftskip-1.7cm
  \includegraphics
  [width=21cm,
  height=5.0cm]
  {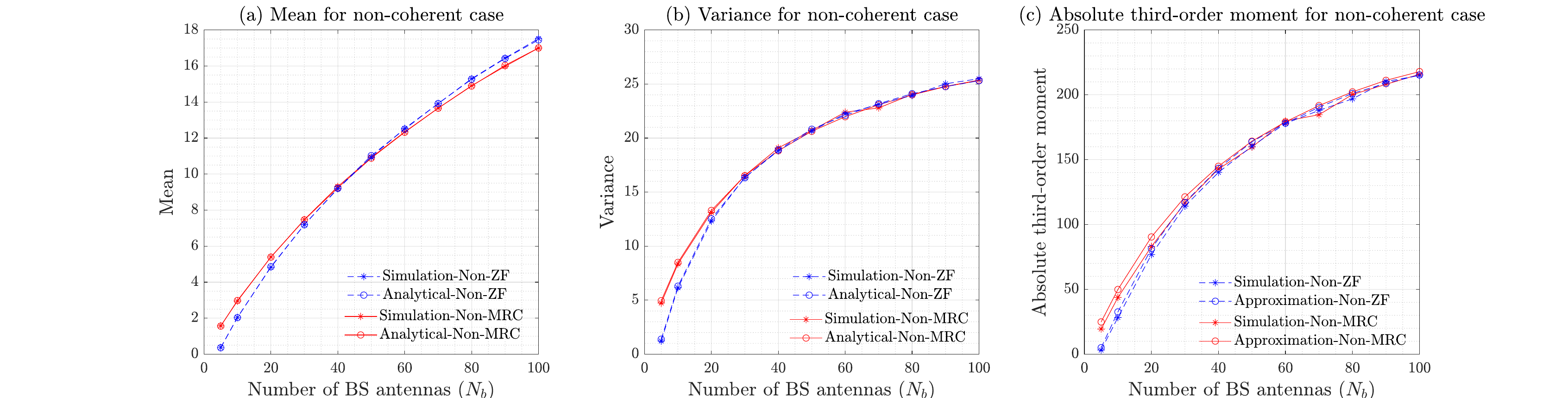}
  \caption{The high-order moments of the information density for the single $j$-th block for both MRC and ZF schemes versus $N_{b}$ with $T_{c}=20$ (non-coherent case).}\label{fig_im_moments}
\end{figure*}

From Appendix \ref{appendix prop_20200103_1} and Appendix \ref{appendix_prop_imper_mrc}, the main results obtained in this paper  rely on the mean, variance and the absolute third-order moment of the corresponding information density. Therefore, Figs. \ref{fig_1}-\ref{fig_im_moments} are used to verify the correctness and accuracy of these results obtained in this paper. Specifically, the correctness of the high-order moments for both MRC and ZF schemes under the coherent case is demonstrated in Fig. \ref{fig_1}, which shows that the simulation and the corresponding analytical results match very well.

Considering the MRC and ZF scheme under the non-coherent case, the closed-form expressions of the mean and variance of the corresponding information density are calculated, and their correctness are verified  in Fig. \ref{fig_im_moments}(a)\&(b). However, for the absolute third-order moment,  it is quite involved and difficult to acquire the exact distributions of the random variables $v^{\text{mrc}}_{y,\text{non}}(j)$ and $v^{\text{mrc}}_{n,\text{non}}(j)$
defined in \eqref{eq_v_yn_mrc} and \eqref{eq_v_yn_mrc-n}, respectively.
Therefore, the matched Gamma distribution approximations are given in \eqref{eq_v_yn_gamma} and \eqref{eq_v_yn_gamma-n}
for the MRC scheme. Likewise, similar random variables and approximations also exist for the ZF scheme. These distribution approximations for both MRC and ZF schemes are simulated in Fig. \ref{fig_im_CDF}.  In this figure, the CDF of the exact distribution and the matched Gamma approximation for  $v^{\text{mrc}}_{y,\text{non}}(j)$ and $v^{\text{mrc}}_{n,\text{non}}(j)$ under the MRC scheme (similar random variables for the ZF scheme)
 are showed. The notation "CDF-Chi-Squ" represents the case when we treat all the effective noise vectors and the corresponding output signal vectors as Gaussian distributed. From this figure, the matched Gamma distribution can precisely evaluate the complicated distributions of the normalized squared norms  and it is not accurate to just simply treat them as Gaussian distributed. Based on these distribution approximations, the absolute third-order moment for both MRC and ZF schemes are then approximated by using Lemma \ref{lemma_20210531_1} in Appendix \ref{preliminary}, and the correctness and accuracy of these approximations are validated in Fig. \ref{fig_im_moments}(c). From this subfigure, we see that the gap between the simulation and the corresponding approximation can be neglected.

Since the analytical expressions agree remarkably well with the corresponding simulations, we will only show the analytical results henceforth unless otherwise stated.
%
\subsection{Perturbation Restrictions}
\begin{figure*}[!t]
  \advance\leftskip-1.7cm
  \includegraphics
  [width=21.0cm,height=4.5cm]
  {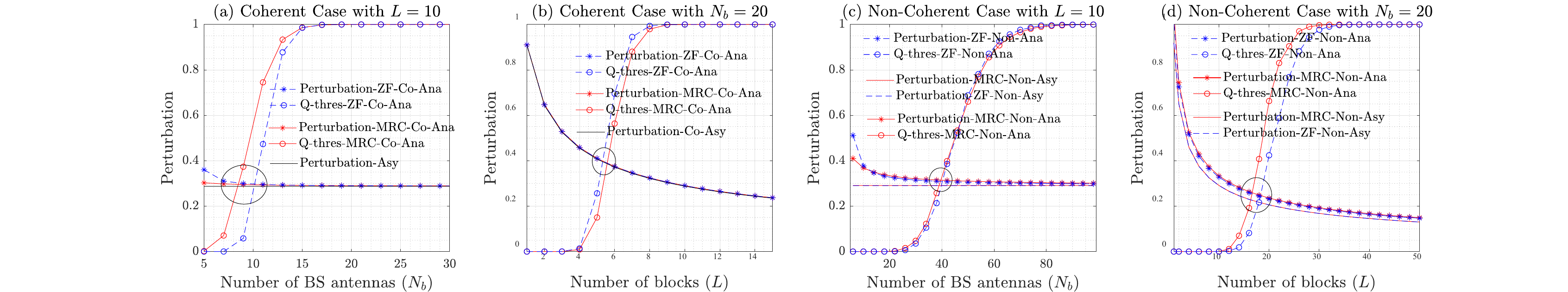}
  \caption{The perturbation and the threshold function for both MRC and ZF schemes.}\label{fig_perturbation_im_co}
\end{figure*}

All the results obtained in this paper are under the condition that the perturbation terms of the achievable coding rate must be restricted below a given threshold.
In Fig. \ref{fig_perturbation_im_co}(a)\&(b) for the coherent case and Fig. \ref{fig_perturbation_im_co}(c)\&(d) for the non-coherent case, the analytical values of the channel perturbations and the corresponding threshold terms are simulated. From Fig. \ref{fig_perturbation_im_co}(a) in the coherent case, when we fix the number of blocks $L$ (e.g., $L=10$, which is a small value), the channel perturbation constraint condition in Propositions \ref{prop_20200103_1}-\ref{prop_imper_mrc} (and the corresponding similar result for ZF scheme in the non-coherent case) can be satisfied quickly by increasing the number of BS antennas (e.g., $N_{b}=10$ in this figure).
However,  increasing the number of BS antennas, the channel perturbation for both MRC and ZF schemes will approach to the same small constant while the threshold terms are approaching to 1. On the contrary, when the number of BS antennas is fixed, Fig. \ref{fig_perturbation_im_co}(b) showcases that only a very small number of blocks ($L=5$) are needed in order to satisfy the channel perturbation constraint
and by increasing the number of blocks, the threshold terms are also approaching to 1 while the perturbation terms approach to zero. Considering the non-coherent case, similar conclusions can be obtained as shown in Fig. \ref{fig_perturbation_im_co}(c)\&(d). The difference from the coherent case of Fig. \ref{fig_perturbation_im_co}(a)\&(b)  is that more BS antennas are needed for the same fixed number of blocks and larger number of blocks are needed for the same fixed number of BS antennas. This is due to the high impact of the channel estimation errors.

Finally, Fig. \ref{fig_perturbation_im_co}  shows that the perturbation terms in this paper are very small even in small-to-moderate number of BS antennas $N_{b}$ and the small number of blocks $L$. This indicates that when massive MIMO is deployed and with even small number of blocks, the channel perturbation of the achievable coding rate for both MRC and ZF schemes under the coherent and non-coherent case can be discarded without affecting the system performance.
\vspace{-3mm}
\subsection{Average Decoding Error Probability}
\begin{figure*}[t!]
  \begin{center}
  \includegraphics
  [
  width=17cm,
  height=6.0cm]
  {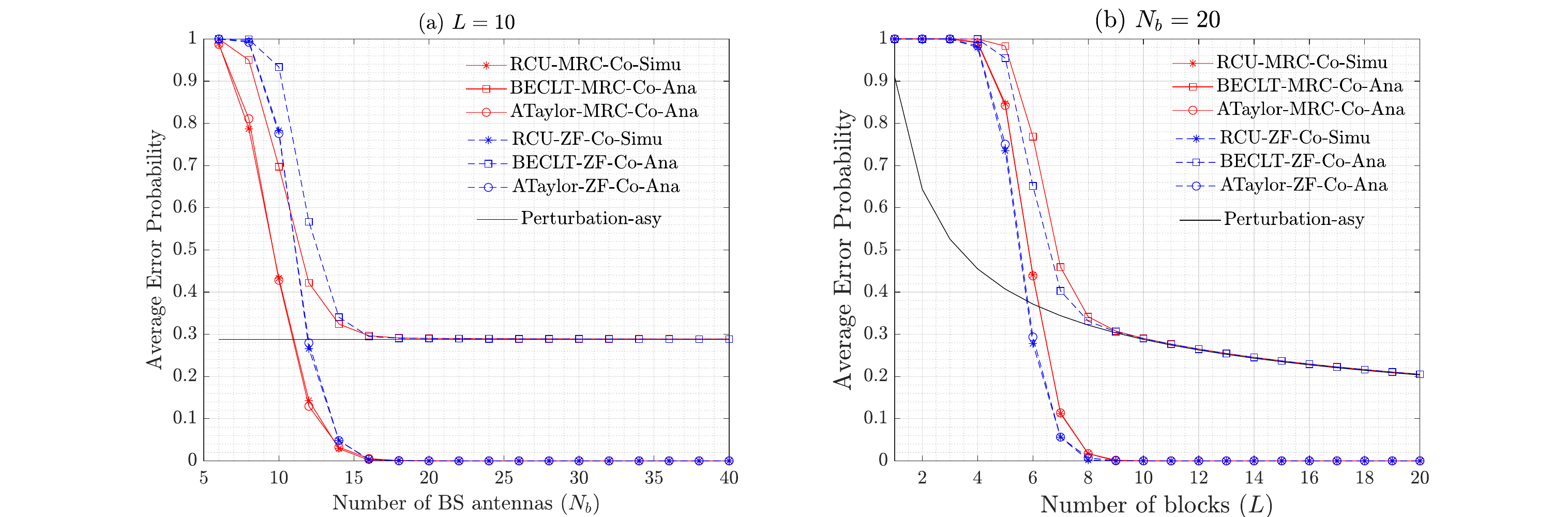}
  \caption{The average error probability from RCU, BECLT, and ATaylor for both MRC and ZF schemes with $b=100$nats,  $P_{k}=10$dBm, and $T_{c}=20$   (coherent case).}\label{fig_3}
  \end{center}
  \vspace{-3mm}
\end{figure*}

\begin{figure*}[t!]
  \advance\leftskip-1.7cm
  \includegraphics
  [width=21cm,
  height=5.0cm]
  {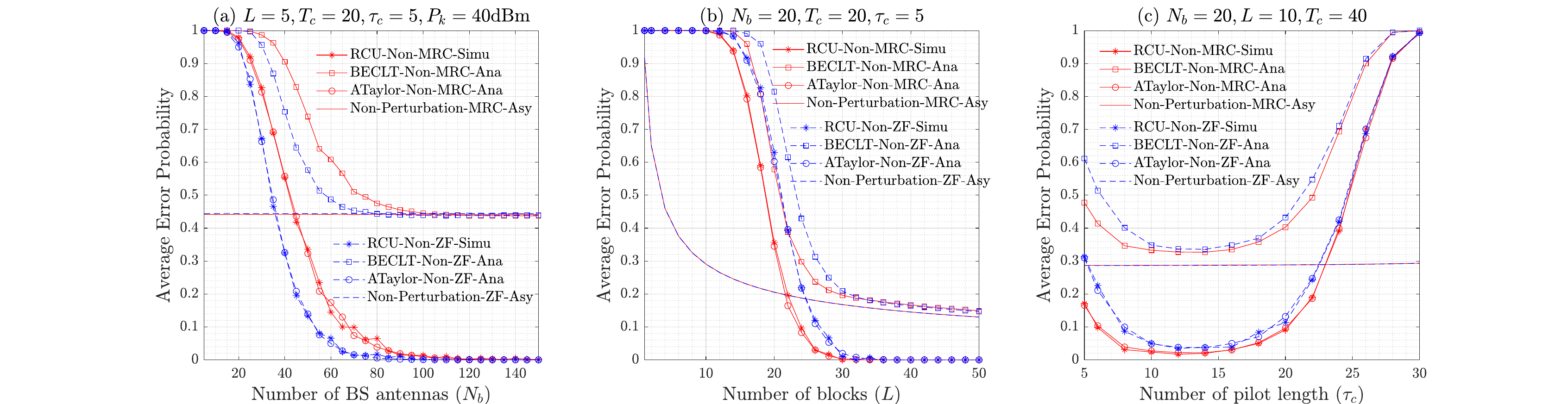}
  \caption{The average error probability from RCU, BECLT, and ATaylor for both MRC and ZF schemes with $b=100$nats and $P_{k}=10$dBm (non-coherent case).
  }\label{fig_AEP_im}
\end{figure*}

All results in this paper are calculated in terms of the RCU bound of the average error probability. Then, the BE-CLT is employed to provide a further upper bound of the average error probability. After this step, when the threshold conditions shown in Fig. \ref{fig_perturbation_im_co} are satisfied, the Taylor expansion is taken to get an approximation of the achievable coding rate, based on  which, an approximation of the average error probability is calculated, which is denoted as "ATaylor" in the figures. In Fig. \ref{fig_3} and Fig. \ref{fig_AEP_im}, the average error probability obtained from these results is illustrated for both MRC and ZF schemes in the coherent and non-coherent case.

Specifically, Fig. \ref{fig_3} shows the simulation result of the RCU bound and the analytical results obtained after the BE-CLT and Taylor approximation  with respect to both $N_{b}$ and $L$ for the coherent case. From this figure, the RCU-based average error probability matches well with the corresponding result obtained from the Taylor approximation, while the BE-CLT-based result acts as an upper bound.\footnote{Note that although the Taylor approximation is more accurate than the BE-CLT-based upper bound, it is not precise enough with an extremely large array at the BS (i.e., the gap from the exact value cannot be ignored) and a more accurate and complicated result was provided in \cite{ostman2021urllc} via the saddlepoint approximations.}
The reason behind this is that when considering multiple blocks, the normalized information density through multiple blocks can be approximated by the normal distribution.
Meanwhile, increasing  the number of BS antennas, the  BE-CLT-based average error probability approaches to a non-zero constant, while the RCU-based and the Taylor-based result  approach to zero. However, by increasing the number of blocks, all the results  approach to zero. The difference between them is that the RCU and Taylor-based results go to zero faster than the result obtained from the BE-CLT. In this context, we can conclude that massive MIMO without the power scaling laws can suppress the average error probability to zero.
Furthermore, the most significant insight extracted from this figure is that it is impractical to compare the performance between the MRC and ZF scheme, which means that in some situations, the MRC scheme performs better than the ZF scheme, as shown in Fig. \ref{fig_3}(a), and in other parameters settings, the ZF scheme has better performance than the MRC scheme, as shown in Fig. \ref{fig_3}(b). This interplay depends on the system parameters, e.g., the number of blocks $L$ and the
${\rm SINR}^{{\text{mrc}}}_{k,\text{co}}$ and ${\rm SNR}^{\text{zf}}_{k,\text{co}}$ through all blocks.

The average error probability for the non-coherent case is shown in Fig. \ref{fig_AEP_im}. Firstly, similar conclusions as extracted from  Fig. \ref{fig_3} for the coherent case can be drawn. The difference is that a higher number of BS antennas and blocks are needed in the non-coherent case.
Then, Fig. \ref{fig_AEP_im}(c)  shows that an optimal value for the  pilot length should be calculated, which is used to make a trade-off between the channel estimation phase and data transmission phase.
\vspace{-3mm}
\subsection{Capacity, Achievable Coding Rate, and Channel Dispersion}
\begin{figure*}[t!]
  \begin{center}
  \includegraphics
  [width=17.0cm,
  height=6.0cm]
  {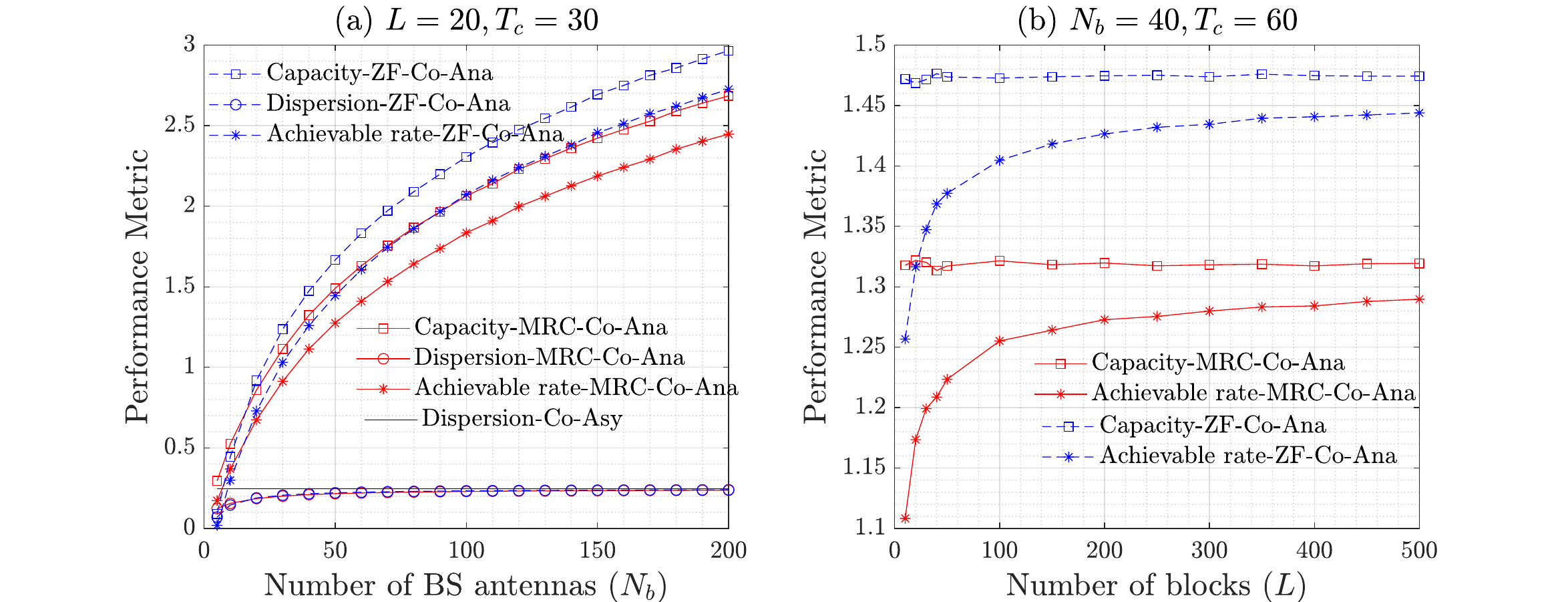}
  \caption{The capacity, achievable coding rate and  channel dispersion for both MRC and ZF schemes with $\epsilon=10^{-5}$ and $P_{k}=10$dBm: (a) versus $N_{b}$  and (b) versus $L$ (coherent case).}\label{fig_4}
  \end{center}
  \vspace{-4mm}
\end{figure*}

\begin{figure*}[t!]
  \advance\leftskip-1.7cm
  \includegraphics
  [width=21.0cm,height=4.5cm]
  {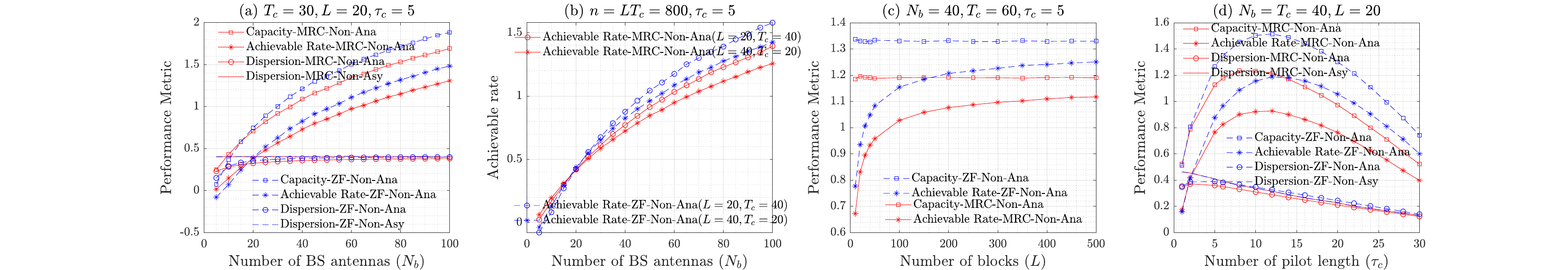}
  \caption{The capacity, achievable coding rate and  channel dispersion for both MRC and ZF schemes with $\epsilon=10^{-5}$ and $P_{k}=30$dBm (non-coherent case).}
  \label{fig_mrc_zf_capacity_rate_dipersion_im}
  \vspace{-3mm}
\end{figure*}

Figures \ref{fig_4}\&\ref{fig_mrc_zf_capacity_rate_dipersion_im} show the
analytical results and the asymptotic results of the channel capacity, the achievable coding rate (after dropping the terms of the order $\mathcal{O}\left(\frac{1}{L}\right)$) and the final effective channel dispersion, i.e., the term ${\small\frac{1}{\sqrt{LT_{c}}}\sqrt{\frac{1}{L}\sum\limits^{L}_{j=1}
{V}_{k}(j)}
Q^{-1}\left(\epsilon_{k}\right)}$ for all situations.
Firstly, Fig. \ref{fig_4} for the coherent case and Fig. \ref{fig_mrc_zf_capacity_rate_dipersion_im} for the non-coherent case illustrate that without considering any power scaling laws, increasing the number of BS antennas can still improve the achievable coding rate without bounds even when the channel dispersion caused by the finite blocklength is taken into account for both MRC and ZF schemes. Meanwhile, the channel dispersion for both  MRC and ZF schemes grows with respect to the number of the BS antennas until an upper bound. Lastly, from Fig. \ref{fig_4}(b) and  Fig. \ref{fig_mrc_zf_capacity_rate_dipersion_im}(c), increasing the number of blocks, the achievable coding rate will grow to its channel capacity, which means that the channel capacity overestimates the performance in the finite blocklength space.

Furthermore, from Fig. \ref{fig_mrc_zf_capacity_rate_dipersion_im}(b) for the non-coherent case, fixing the total number of blocklength $n=LT_{c}=800$, the scenarios with longer coherence intervals and smaller number of blocks can offer higher achievable coding rate for both MRC and ZF schemes. This is due to that under this circumstance, we do not need to estimate the channels frequently, such that more resources could be allocated for the data transmission. At the same time, from Fig. \ref{fig_mrc_zf_capacity_rate_dipersion_im}(d), we have that increasing the pilot length $\tau_{c}$, the channel dispersion for both MRC and ZF schemes are almost diminishing and an optimal value of the pilot length $\tau_{c}$ should be calculated for both the achievable coding rate and the channel capacity.
On this optimal point, the gap between the achievable coding rate and the corresponding capacity is the largest for both MRC and ZF schemes.
\vspace{-3mm}
\subsection{Packet Loss Probability}
\begin{figure*}[t!]
  \begin{center}
  \includegraphics
  [width=17.0cm,
  height=6.0cm]{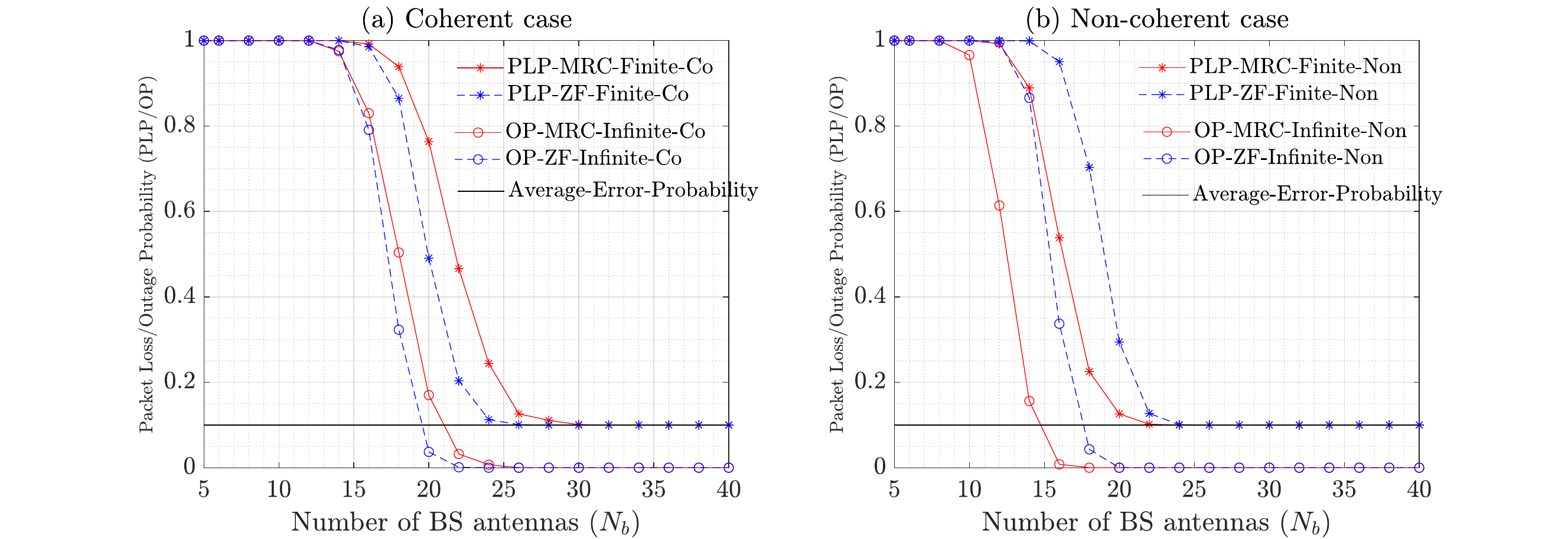}
  \caption{The packet loss probability (PLP) and outage probability (OP) in the finite and infinite blocklength regime respectively
  for both MRC and ZF schemes
  versus $N_{b}$: (a) coherent case with $L=10$, $T_{c}=20$, and $\tilde{R}^{\text{thres}}_{k}=0.8$; (b) non-coherent case with $L=20$, $T_{c}=30$, and $\tilde{R}^{\text{thres}}_{k}=0.2$.}\label{fig_PLP}
  \end{center}
  \vspace{-4mm}
\end{figure*}

Figure \ref{fig_PLP} shows the variation of the packet loss probability given in Corollary \ref{theorem-2021-11-10-1} against the number of BS antennas, and at the same time  we include curves of the outage probability in the infinite blocklength regime. Note that to exhibit the results clearly, we set a larger average error probability, i.e.,  $\epsilon_{k}=0.1$. From this figure, the packet loss probability will approach to the presented average error probability while the outage probability in the infinite blocklength regime tends to be zero with an increasing number of the BS antennas.
This indicates that the average error probability caused by the finite blocklengh is the basic source of a packet loss.
\vspace{-3mm}
\section{Conclusion}
\label{conclusion}
We have investigated  an  MU-MIMO uplink system in the finite blocklength regime over both coherent and non-coherent  block-fading channels. Both the MRC and ZF combining schemes at the BS were considered. The matched ML decoding metric was deployed in the coherent case while the mismatched NN decoding metric was applied in the non-coherent case.
The input signals were assumed to be i.i.d. Gaussian sequences. Under this system setting, we derived closed-form results for the coherent case and tight approximations for the  non-coherent case of the channel perturbation terms of the achievable coding rate, while exact results for the channel dispersions of the achievable coding rate under all cases were also presented.
After providing a constraint threshold of the channel perturbations, the relation between the achievable coding rate and the corresponding average error probability were then derived. Based on these analytical results, an asymptotic analysis for a massive antenna array at the BS with and without the power scaling laws was performed and a series of useful insights were extracted.
\appendices
\vspace{-3mm}
\section{Preliminaries}\label{preliminary}
\begin{lemma}\label{lemma_variance_gamma}
\cite{gaunt2014variance} For two correlated chi-square distributed random variables $x$ and $y$ with the distribution  $x,y\sim \chi^{2}(N)$ and the correlation $\rho$, then, the random variable $z\triangleq x-y$ is variance-gamma distributed with the corresponding PDF  given by
%
\begin{align}\label{eq-variance-gamma}
f_{Z}(z)&=\frac{1}{2^{N} \sqrt{\pi}}\frac{1}{\sqrt{1-\rho}} \frac{1}{\Gamma\left(\frac{N}{2}\right)}
\left|\frac{z}{\sqrt{1-\rho}}\right|^{\frac{N-1}{2}}\notag\\
&\quad\times K_{\frac{N-1}{2}}\left(\frac{1}{2\sqrt{1-\rho}}|z|\right),
\end{align}
where $K_{a}(\cdot)$ denotes the modified Bessel functions  of the second kind.
\end{lemma}
\begin{lemma}\label{lemma_20210531_1}
Consider any two correlated Gamma distributed random variables with the  distributions $x\sim \Gamma(a_{1},b_{1})$ and $y\sim \Gamma(a_{2},b_{2})$ and the correlation coefficient $\rho$. Define the index
\begin{align}
&
i_{\max}\triangleq \left\{i_{\max}|a_{i_{\max}}=\max\{a_{1},a_{2}\}\right\},\\
\quad
&
i_{\min}\triangleq i_{\max}+(-1)^{i_{\max}+1},
\end{align}
and the coefficients in \eqref{eq-2022-08-10-1} and \eqref{eq-2022-08-10-2} shown at the top of next page.
%
\begin{figure*}[!ht]
\begin{align}\label{eq-2022-08-10-1}
l_{11}=\frac{\sqrt{a_{i_{\min}}}}{b_{i_{\min}}},
\quad l_{21}=\rho\frac{\sqrt{a_{i_{\max}}}}{b_{i_{\max}}},
\quad l_{22}=\sqrt{1-\rho^2}\frac{\sqrt{a_{i_{\max}}}}{b_{i_{\max}}}.
\end{align}
\begin{align}\label{eq-2022-08-10-2}
\alpha^{w}_{1}=a_{i_{\min}},
\quad\beta^{w}_{1}=\sqrt{a_{i_{\min}}},
\quad\alpha^{w}_{2}=\frac{(\sqrt{a_{i_{\max}}}
-\rho\sqrt{a_{i_{\min}}})^2}{1-\rho^2},
\quad \beta^{w}_{2}=\frac{\sqrt{a_{i_{\max}}}
-\rho\sqrt{a_{i_{\min}}}}{\sqrt{1-\rho^2}}.
\end{align}
\center\hrulefill
\vspace{-2mm}
\end{figure*}
Then, the  absolute high-order moment of the random variable $z=x-y$ is as follows:

(I) When $l_{21}\geq l_{11}$, i.e., the correlation coefficient $\rho\geq \frac{\sqrt{a_{i_{\min}}}}{\sqrt{a_{i_{\max}}}}
\frac{b_{i_{\max}}}{b_{i_{\min}}}$,
 we have
%
\begin{align}
\mathbb{E}\left\{|x-y|^{m}\right\}
&\approx\sum\limits^{m}_{f=0}\frac{m!}{f!(m-f)!}
\left(\frac{l_{21}-l_{11}}{\beta^{w}_{1}}\right)^{f}
\left(\frac{l_{22}}{\beta^{w}_{2}}\right)^{m-f}\notag\\
&\quad\times\frac{\Gamma(f+\alpha^{w}_{1})}
{\Gamma(\alpha^{w}_{1})}
\frac{\Gamma(m-f+\alpha^{w}_{2})}
{\Gamma(\alpha^{w}_{2})}.
\end{align}

(II) When $l_{21}< l_{11}$, i.e., the correlation coefficient $0\leq \rho\leq \frac{\sqrt{a_{i_{\min}}}}{\sqrt{a_{i_{\max}}}}
\frac{b_{i_{\max}}}{b_{i_{\min}}}$, we have \eqref{eq-2022-08-10-3} shown at the top of next page.
%
\begin{figure*}[!ht]
\begin{align}\label{eq-2022-08-10-3}
&\quad\mathbb{E}\left\{|x-y|^{m}\right\}
\approx\frac{(\beta^{w}_{1})^{\alpha^{w}_{1}}
(\beta^{w}_{2})^{\alpha^{w}_{2}}
l^{m+\alpha^{w}_{1}}_{22}
(l_{11}-l_{21})^{m+\alpha^{w}_{2}}}
{[l_{22}\beta^{w}_{1}
+(l_{11}-l_{21})\beta^{w}_{2}
]^{m+\alpha^{w}_{2}+\alpha^{w}_{1}}}
\frac{\Gamma\left(m+\alpha^{w}_{2}+\alpha^{w}_{1}\right)}
{\Gamma(\alpha^{w}_{1})\Gamma(\alpha^{w}_{2})}
\sum\limits^{m}_{f=0}\frac{(-1)^{f}m!}{f!(m-f)!}
\notag\\
&
\times\!
\left\{\!
\frac{{}_{2}F_{1}\!\left(1,m\!+\!\alpha^{w}_{2}
\!+\!\alpha^{w}_{1},
f\!+\!\alpha^{w}_{1}\!+\!1;
\frac{l_{22}\beta^{w}_{1}}{l_{22}
\beta^{w}_{1}\!+\!(l_{11}\!-\!l_{21})
\beta^{w}_{2}}\right)}{f\!+\!\alpha^{w}_{1}}
\!+\!\frac{{}_{2}F_{1}\!\left(1,m\!+\!\alpha^{w}_{2}
\!+\!\alpha^{w}_{1},
f\!+\!\alpha^{w}_{2}\!+\!1;
\frac{(l_{11}\!-\!l_{21})\beta^{w}_{2}}{l_{22}
\beta^{w}_{1}\!+\!(l_{11}\!-\!l_{21})\beta^{w}_{2}}
\right)}{f\!+\!\alpha^{w}_{2}}\!
\right\}.
\end{align}
\center\hrulefill
\vspace{-3mm}
\end{figure*}
\end{lemma}
\vspace{2mm}
\begin{IEEEproof}
The key idea is to use the
Cholesky decomposition to transform correlated Gamma distributed random variables into a linear combination of independent Gamma distributed random variables, as given in \cite{zhang2004simulation}. The detailed proof of Lemma \ref{lemma_20210531_1} is omitted due to limited space.
\end{IEEEproof}
%
\section{Proof of Proposition \ref{prop_20200103_1}}
\label{appendix prop_20200103_1}
Our proof resembles
those developed by Polyanskiy \textit{et al}. in \cite{polyanskiy2010channel}. So, we
begin by presenting the definition of the generalized information density, which is given by \cite{lancho2020saddlepoint}\footnote{Note that our proof is conditioned on the uplink channels $\{{\bf h}_{k}(j)\}^{j=1,\ldots,L}_{k=1,\ldots,K_{u}}$ and for simplicity, we removed this condition from the entire proof.}
%
\begin{align}\label{eq_per_mrc_20210316_1}
&\quad
i_{s}\left(({\bf X}_{k})^{L};
\left((\tilde{{\bf Y}}_{k})_{\text{co}}^{\text{mrc}}\right)^{L}\right)\notag\\
&\!=\!\ln \frac{\operatorname{p}^{s}\left\{\left.\left((\tilde{{\bf Y}}_{k})_{\text{co}}^{\text{mrc}}\right)^{L}\right|
\left({\bf X}_{k}\right)^{L}\right\}}{\mathbb{E}_{(\bar{{\bf X}}_{k})^{L}}\left\{
\operatorname{p}^{s}\left\{\left.\left((\tilde{{\bf Y}}_{k})_{\text{co}}^{\text{mrc}}\right)^{L}\right|
\left(\bar{{\bf X}}_{k}\right)^{L}\right\}\right\}}\notag\\
&\!=\!\sum\limits^{L}_{j=1}i_{s}\left({\bf x}_{k}(j); \tilde{{\bf y}}^{\text{mrc}}_{k,\text{co}}(j)\right), \quad s>0,
\end{align}
%
where ${\small\operatorname{p}^{s}\left\{\left.\left((\tilde{{\bf Y}}_{k})_{\text{co}}^{\text{mrc}}\right)^{L}\right|
\left({\bf X}_{k}\right)^{L}\right\}}$ has been given in \eqref{eq_20200731_1}. The notation $(\bar{{\bf X}}_{k})^{L}$ is defined as $(\bar{{\bf X}}_{k})^{L}\triangleq [\bar{{\bf x}}_{k}(1),\ldots,\bar{{\bf x}}_{k}(L)]$, in which $\bar{{\bf x}}_{k}(j)\in\mathcal{C}^{T_{c}\times 1}$ is a random vector that is independent with ${\bf x}_{k}(j)$  and has the same distribution as  ${\bf x}_{k}(j)$, i.e., $\bar{{\bf x}}_{k}(j)\sim\mathcal{CN}(0, P_{k}{\bf I}_{T_{c}})$.
The term $i_{s}\left({\bf x}_{k}(j); \tilde{{\bf y}}^{\text{mrc}}_{k,\text{co}}(j)\right)$ denotes the generalized information density at the $j$-th block,  given by
%
\begin{align}\label{eq_per_mrc_20210319_7}
&\quad i_{s}\left({\bf x}_{k}(j); \tilde{{\bf y}}^{\text{mrc}}_{k,\text{co}}(j)\right)\notag\\
&\!=\!T_{c}\ln\left(1\!+\!s\frac{P_{k}\|{\bf h}_{k}(j)\|^{2}_{2}}{\sigma^{2}_{\check{{\bf n}}_{\text{co}}^{\text{mrc}}}(j)}\right)\notag\\
&\quad\!+\!\frac{s\|\tilde{{\bf y}}^{\text{mrc}}_{k,\text{co}}(j)\|^{2}_{2}}
{\sigma^{2}_{\check{{\bf n}}_{\text{co}}^{\text{mrc}}}(j)\!+\!sP_{k}\|{\bf h}_{k}(j)\|^{2}_{2}}
\!-\!\frac{s\|\check{{\bf n}}^{\text{mrc}}_{\text{co}}(j)\|^{2}_{2}}
{\sigma^{2}_{\check{{\bf n}}_{\text{co}}^{\text{mrc}}}(j)},
\end{align}
which shows that conditioned on the uplink channels at the BS, the distribution of information density is only determined by the squared norm of the output signal vector and the  effective noise vector, i.e., $\|\tilde{{\bf y}}^{\text{mrc}}_{k,\text{co}}(j)\|^{2}_{2}$ and $\|\check{{\bf n}}^{\text{mrc}}_{\text{co}}(j)\|^{2}_{2}$.

Then, with the random coding union bound in \cite[Theorem 1]{martinez2011saddlepoint}, it is easy to obtain that the  average decoding error probability
can be upper bounded by
%
\begin{align}\label{eq_20200115_1}
\!\!\epsilon^{\text{mrc}}_{k,\text{co}}
\!\leq \!
\operatorname{Pr}\!\left\{\!i_{s}\left(\!({\bf X}_{k})^{L};
\left((\tilde{{\bf Y}}_{k})_{\text{co}}^{\text{mrc}}\!\right)^{L}\right)
%
\!\leq \!\ln(M_{k}-1)\!-\!\ln \mu\!\right\},
\end{align}
%
where $\mu$ is a random variable that is uniformly distributed on the interval $(0,1)$.

Next, conditioned on the CSI at the BS,
we focus on the sequence  $\left\{i_{s}\left({\bf x}_{k}(j); \tilde{{\bf y}}^{\text{mrc}}_{k,\text{co}}(j)
\right), j=1,\ldots,L\right\}$ consisting of independent random variables with their mean, variance and absolute third-order moment being $\tilde{I}_{k,\text{co}}^{\text{mrc}}(j)$, $\tilde{V}_{k,\text{co}}^{\text{mrc}}(j)$ and
$\tilde{U}_{k,\text{co}}^{\text{mrc}}(j)$,
respectively. Then, \eqref{eq_20200115_1} can be rewritten as \eqref{eq_20200731_4} shown at the top of next page,
%
\begin{figure*}[!ht]
\begin{align}\label{eq_20200731_4}
\epsilon^{\text{mrc}}_{k,\text{co}}
&\leq
\operatorname{Pr}\left\{\frac{
\sum\limits^{L}_{j=1}\left[i_{s}\left({\bf x}_{k}(j); \tilde{{\bf y}}^{\text{mrc}}_{k,\text{co}}(j)\right)
-\tilde{I}_{k,\text{co}}^{\text{mrc}}(j)\right]}
{\sqrt{\sum\limits^{L}_{j=1}
\tilde{V}_{k,\text{co}}^{\text{mrc}}(j)}}
\leq
\frac{\ln (M_{k}\!-\!1)-\ln\mu\!-\!\sum\limits^{L}_{j=1}
\tilde{I}_{k,\text{co}}^{\text{mrc}}(j)}
{\sqrt{\sum\limits^{L}_{j=1}
\tilde{V}_{k,\text{co}}^{\text{mrc}}(j)}}\right\}\notag\\
&\overset{(a)}{\leq}\min\left\{1,
\left[1\!-\!Q\left(\frac{\ln (M_{k}-1)-\ln\mu-\sum\limits^{L}_{j=1}
\tilde{I}_{k,\text{co}}^{\text{mrc}}(j)}
{\sqrt{\sum\limits^{L}_{j=1}
\tilde{V}_{k,\text{co}}^{\text{mrc}}(j)}}\right)
+\frac{c_{1}\sum\limits^{L}_{j=1} \tilde{U}_{k,\text{co}}^{\text{mrc}}(j)}
{\left(\sum\limits^{L}_{j=1}
\tilde{V}_{k,\text{co}}^{\text{mrc}}(j)
\right)^{\frac{3}{2}}}\right]\right\},
\end{align}
\center\hrulefill
\end{figure*}
where (a) is due to the Berry-Esseen central limit theorem (BE-CLT) \cite{maya2016closed} and the coefficient $c_{1}=0.56$. In \eqref{eq_20200731_4}, we define the term
%
\begin{align}\label{eq_per_mrc_20210317_1}
{U}^{\text{mrc}}_{k,\text{co}}
\triangleq
c_{1}\left(\sum\limits^{L}_{j=1}
\tilde{V}_{k,\text{co}}^{\text{mrc}}(j)\right)^{-\frac{3}{2}}
\left(\sum\limits^{L}_{j=1} \tilde{U}_{k,\text{co}}^{\text{mrc}}(j)\right),
\end{align}
\begin{align}\label{eq_per_mrc_20210317_1-1}
\tilde{q}^{\text{mrc},\text{co}}_{k,\text{thres}}
&\triangleq
Q\left(\left(\sum\limits^{L}_{j=1}
\tilde{V}_{k,\text{co}}^{\text{mrc}}(j)\right)^{-\frac{1}{2}}
\right.\notag\\
&\left.\quad\times
\left(\ln (M_{k}-1)-\ln\mu-\sum\limits^{L}_{j=1}
\tilde{I}_{k,\text{co}}^{\text{mrc}}(j)\right)
\right),
\end{align}
%
then, the results \eqref{eq_20210923_2} and \eqref{eq_20210305_2} of Proposition \ref{prop_20200103_1} can be obtained by first determining the exact value of ${U}^{\text{mrc}}_{k,\text{co}}$ and $\tilde{q}^{\text{mrc},\text{co}}_{k,\text{thres}}$
and then taking the Taylor expansion on the function $Q^{-1}\left(1-{\epsilon}_{k,\text{co}}^{\text{mrc}}
+{U}^{\text{mrc}}_{k,\text{co}}\right)$ at the point $1-{\epsilon}_{k,\text{co}}^{\text{mrc}}$.
Now, from \eqref{eq_20200731_4}-\eqref{eq_per_mrc_20210317_1-1}, we need to calculate the mean, variance, and absolute third-order moment of the generalized information density of \eqref{eq_per_mrc_20210319_7}.

Under the condition on the uplink channels $\{{\bf h}_{k}(j)\}^{j=1,\ldots,L}_{k=1,\ldots,K_{u}}$ at the BS,
the mean value is
%
\begin{align}\label{eq_per_mrc_20210319_8}
\tilde{I}_{k,\text{co}}^{\text{mrc}}(j)
&=\mathbb{E}\left\{i_{s}\left({\bf x}_{k}(j); \tilde{{\bf y}}^{\text{mrc}}_{k,\text{co}}(j)\right)\right\}
\notag\\
&
=T_{c}\left\{\ln\left(1+s\frac{P_{k}\|{\bf h}_{k}(j)\|^{2}_{2}}{\sigma^{2}_{\check{{\bf n}}_{\text{co}}^{\text{mrc}}}(j)}\right)\right.\notag\\
&\left.\quad+\frac{s\left(\sigma^{2}_{\check{{\bf n}}_{\text{co}}^{\text{mrc}}}(j)+P_{k}\|{\bf h}_{k}(j)\|^{2}_{2}\right)}{\sigma^{2}_{\check{{\bf n}}_{\text{co}}^{\text{mrc}}}(j)+sP_{k}\|{\bf h}_{k}(j)\|^{2}_{2}}
-s\right\}.
\end{align}
%
Then, the optimal solution of the optimization problem ${\small \sup\limits_{s>0} \tilde{I}_{k,\text{co}}^{\text{mrc}}(j)}$ is $s=1$, under which, the channel capacity is achieved.
Thus, in the following, the variance and the absolute third-order  moment are derived under the choice of $s=1$ in \eqref{eq_per_mrc_20210319_7} and \eqref{eq_per_mrc_20210319_8}, which are given by \eqref{eq_per_mrc_20210319_9} shown at the top of next page.
%
\begin{figure*}[!ht]
\begin{align}\label{eq_per_mrc_20210319_9}
&\quad
\mathbb{E}\left\{\left|i_{s=1}\left({\bf x}_{k}(j); \tilde{{\bf y}}^{\text{mrc}}_{k,\text{co}}(j)\right)
-\mathbb{E}\left\{i_{s=1}\left({\bf x}_{k}(j); \tilde{{\bf y}}^{\text{mrc}}_{k,\text{co}}(j)\right)\right\}
\right|^{m}\right\}
\notag\\
&=\frac{1}{2^{m}}
\mathbb{E}\left\{\left|\frac{2\|\tilde{{\bf y}}^{\text{mrc}}_{k,\text{co}}(j)\|^{2}_{2}}{\sigma^{2}_{\check{{\bf n}}_{\text{co}}^{\text{mrc}}}(j)+P_{k}\|{\bf h}_{k}(j)\|^{2}_{2}}
-\frac{2\|\check{{\bf n}}^{\text{mrc}}_{\text{co}}(j)\|^{2}_{2}}{\sigma^{2}_{\check{{\bf n}}_{\text{co}}^{\text{mrc}}}(j)}\right|^{m}\right\},
\quad m=2,3.
\end{align}
\center\hrulefill
\vspace{-3mm}
\end{figure*}
%
In \eqref{eq_per_mrc_20210319_9}, both the random variables $\frac{2\left\|\tilde{{\bf y}}^{\text{mrc}}_{k,\text{co}}(j)\right\|^{2}_{2}}{\sigma^{2}_{\check{{\bf n}}_{\text{co}}^{\text{mrc}}}(j)+P_{k}\|{\bf h}_{k}(j)\|^{2}_{2}}$ and $\frac{2\left\|\check{{\bf n}}^{\text{mrc}}_{\text{co}}(j)\right\|^{2}_{2}}{\sigma^{2}_{\check{{\bf n}}_{\text{co}}^{\text{mrc}}}(j)}$ are chi-square distributed with the distribution
$\chi^{2}(2T_{c})$  and  the correlation coefficient $\rho^{\text{mrc}}_{\text{co}}$ given by
%
\begin{align}\label{eq_per_mrc_20210319_10}
\rho^{\text{mrc}}_{\text{co}}
&=\frac{1}{4T_{c}}
\mathbb{E}\left\{\left[\frac{2\left\|\tilde{{\bf y}}^{\text{mrc}}_{k,\text{co}}(j)\right\|^{2}_{2}}{\sigma^{2}_{\check{{\bf n}}_{\text{co}}^{\text{mrc}}}(j)+P_{k}\|{\bf h}_{k}(j)\|^{2}_{2}}-2T_{c}\right]\right.\notag\\
&\quad\left.\times\left[\frac{2\left\|\check{{\bf n}}^{\text{mrc}}_{\text{co}}(j)\right\|^{2}_{2}}
{\sigma^{2}_{\check{{\bf n}}_{\text{co}}^{\text{mrc}}}(j)}-2T_{c}\right]
\right\}
\notag\\
&
=\frac{\sigma^{2}_{\check{{\bf n}}_{\text{co}}^{\text{mrc}}}(j)}{P_{k}\|{\bf h}_{k}(j)\|^{2}_{2}
+\sigma^{2}_{\check{{\bf n}}_{\text{co}}^{\text{mrc}}}(j)}.
\end{align}
%
Finally,  the variance $\tilde{V}_{k,\text{co}}^{\text{mrc}}(j)$ and the absolute third-order moment $\tilde{U}_{k,\text{co}}^{\text{mrc}}(j)$ of the information density $i_{s=1}\left({\bf x}_{k}(j); \tilde{{\bf y}}^{\text{mrc}}_{k,\text{co}}(j)\right)$ can be obtained from \eqref{eq_per_mrc_20210319_9} by using Lemma \ref{lemma_variance_gamma} and setting $m=2$ and $m=3$, respectively.
\vspace{-3mm}
\section{Proof of Proposition \ref{prop_imper_mrc}}
\label{appendix_prop_imper_mrc}
Similar as Appendix \ref{appendix prop_20200103_1}, our proof also begins by the definition of the generalized information density, which is given similarly as \eqref{eq_per_mrc_20210316_1} by replacing ${\small\operatorname{p}^{s}\left\{\left.\left((\tilde{{\bf Y}}_{k})_{\text{co}}^{\text{mrc}}\right)^{L}\right|
\left({\bf X}_{k}\right)^{L}\right\}}$ with
${\small\exp\left\{-s\left\|((\tilde{{\bf Y}}_{k})_{\text{non}}^{\text{mrc}})^{L}-(\hat{{\bf H}}_{k})^{L}(({\bf X}_{k})^{L})^{T}\right\|^{2}_{F}\right\}}$ given in \eqref{eq_p_zf_20210307_1} and the corresponding expression is
%
\begin{align}\label{eq_per_mrc_20210319_12}
i_{s}\left(({\bf X}_{k})^{L};
((\tilde{{\bf Y}}_{k})_{\text{non}}^{\text{mrc}})^{L}\right)
=\sum\limits^{L}_{j=1}i_{s}\left({\bf x}_{k}(j); \tilde{{\bf y}}^{\text{mrc}}_{k,\text{non}}(j)\right),
\end{align}
%
where the generalized information density  $i_{s}\left({\bf x}_{k}(j); \tilde{{\bf y}}^{\text{mrc}}_{k,\text{non}}(j)\right)$ at the $j$-th block is given by
%
\begin{align}\label{eq_per_mrc_20210319_13}
&\quad i_{s}\left({\bf x}_{k}(j); \tilde{{\bf y}}^{\text{mrc}}_{k,\text{non}}(j)\right)\notag\\
&\!=\!(T_{c}\!-\!\tau_{c})\ln\left(1\!+\!sP_{k}\|\hat{{\bf h}}^{\dagger}_{k}(j)\|^{2}_{2}\right)\notag\\
&\quad\!+\!\frac{s\left\|\tilde{{\bf y}}^{\text{mrc}}_{k,\text{non}}(j)\right\|_{2}^{2}}
{1\!+\!sP_{k}\|\hat{{\bf h}}^{\dagger}_{k}(j)\|^{2}_{2}}
\!-\!s\|\check{{\bf n}}^{\text{mrc}}_{\text{non}}(j)\|^{2}_{2}.
\end{align}
%
Then, the proof can be obtained by following the similar process as given in Appendix \ref{appendix prop_20200103_1}. Hence, we only need to calculate the mean $\tilde{I}_{k,\text{non}}^{\text{mrc}}(j)$, variance $\tilde{V}_{k,\text{non}}^{\text{mrc}}(j)$ and the absolute third-order  moment $\tilde{U}_{k,\text{non}}^{\text{mrc}}(j)$ of the generalized information density $i_{s}\left({\bf x}_{k}(j); \tilde{{\bf y}}^{\text{mrc}}_{k,\text{non}}(j)\right)$ of \eqref{eq_per_mrc_20210319_13}.

Under the condition of the uplink estimated channels $\{\hat{{\bf h}}_{k}(j)\}^{j=1,\ldots,L}_{k=1,\ldots,K_{u}}$ at the BS, the mean value $\tilde{I}_{k,\text{non}}^{\text{mrc}}(j)
=\mathbb{E}\left\{i_{s}\left({\bf x}_{k}(j); \tilde{{\bf y}}^{\text{mrc}}_{k,\text{non}}(j)\right)\right\}$ is
%
\begin{align}\label{eq_per_mrc_20210319_14}
\tilde{I}_{k,\text{non}}^{\text{mrc}}(j)
&=(T_{c}-\tau_{c})\left\{\ln\left(1+sP_{k}\|\hat{{\bf h}}^{\dagger}_{k}(j)\|^{2}_{2}\right)\right.\notag\\
&\left.\quad+\frac{sP_{k}\|\hat{{\bf h}}^{\dagger}_{k}(j)\|^{2}_{2}(1-s\sigma^{2}_{\check{{\bf n}}^{\text{mrc}}_{\text{non}}}(j))}{1+sP_{k}\|\hat{{\bf h}}^{\dagger}_{k}(j)\|^{2}_{2}}\right\}.
\end{align}
Then, the solution of the following optimization problem with respect to $s>0$
%
\begin{align}\label{eq_optimization_s}
\sup_{s>0} \tilde{I}_{k,\text{non}}^{\text{mrc}}(j),
\end{align}
%
is that $s=\frac{1}{\sigma^{2}_{\check{{\bf n}}^{\text{mrc}}_{\text{non}}}(j)}$,
which is consistent with the result in \cite{scarlett2020information}.
With this given $s$, the channel capacity for the mismatched decoding metric under the non-coherent case can be achieved.

Thus, in the following, the variance $\tilde{V}_{k,\text{non}}^{\text{mrc}}(j)$ and the absolute third-order moment $\tilde{U}_{k,\text{non}}^{\text{mrc}}(j)$ of \eqref{eq_per_mrc_20210319_13} will be derived under the setting $s=\frac{1}{\sigma^{2}_{\check{{\bf n}}^{\text{mrc}}_{\text{non}}}(j)}$, which are equivalent to derive
%
\begin{align}\label{eq_per_mrc_20210319_15}
&\quad\mathbb{E}\left\{\left|\frac{\|\tilde{{\bf y}}^{\text{mrc}}_{k,\text{non}}(j)\|_{2}^{2}}
{P_{k}\|\hat{{\bf h}}^{\dagger}_{k}(j)\|^{2}_{2}+\sigma^{2}_{\check{{\bf n}}^{\text{mrc}}_{\text{non}}}(j)}
-\frac{\|\check{{\bf n}}^{\text{mrc}}_{\text{non}}(j)\|^{2}_{2}}{\sigma^{2}_{\check{{\bf n}}^{\text{mrc}}_{\text{non}}}(j)}\right|^{m}\right\}\notag\\
&\!=\!\mathbb{E}\left\{\left|v^{\text{mrc}}_{y,\text{non}}(j)
-v^{\text{mrc}}_{n,\text{non}}(j)
\right|^{m}\right\}
, \!\quad m=2,3,
\end{align}
%
where $v^{\text{mrc}}_{y,\text{non}}(j)$ and $v^{\text{mrc}}_{n,\text{non}}(j)$ have been defined in \eqref{eq_v_yn_mrc} and \eqref{eq_v_yn_mrc-n}, respectively.
Substituting \eqref{eq_im_20210307_2} and \eqref{eq_im_20210307_1}  of Section \ref{section_imperfect} into \eqref{eq_per_mrc_20210319_15},  we can get that the closed-form result of the variance is $\tilde{V}_{k,\text{non}}^{\text{mrc}}(j)=T_{c}
{V}_{k,\text{non}}^{\text{mrc}}(j)$, in which ${V}_{k,\text{non}}^{\text{mrc}}(j)$ is as given in \eqref{eq_20210311_5}.
However, for the calculation of the absolute third-order moment
$\tilde{U}_{k,\text{non}}^{\text{mrc}}(j)$, i.e., for $m=3$ in \eqref{eq_per_mrc_20210319_15}.
Generally speaking,  the calculation of $\tilde{U}_{k,\text{non}}^{\text{mrc}}(j)$ by using the exact distributions of $v^{\text{mrc}}_{y,\text{non}}(j)$ and $v^{\text{mrc}}_{n,\text{non}}(j)$
is quite involved due to the difficulty and computational complexity of deriving the exact PDF of them.
Thus, we invoke the matched Gamma distribution approximations of $v^{\text{mrc}}_{y,\text{non}}(j)$ and $v^{\text{mrc}}_{n,\text{non}}(j)$ as given in \eqref{eq_v_yn_gamma} and \eqref{eq_v_yn_gamma-n}
to derive $\tilde{U}_{k,\text{non}}^{\text{mrc}}(j)$, which is  obtained as given in \eqref{eq_20210311_7} by directly using the result in Lemma \ref{lemma_20210531_1}.
%
\vspace{-2mm}
%
\def\baselinestretch{0.90}
\end{document}